\documentclass[fleqn,usenatbib]{mnras}

\usepackage{newtxtext,newtxmath}

\usepackage[T1]{fontenc}

\DeclareRobustCommand{\VAN}[3]{#2}
\let\VANthebibliography\thebibliography
\def\thebibliography{\DeclareRobustCommand{\VAN}[3]{##3}\VANthebibliography}

\usepackage{graphicx}	
\usepackage{amsmath}	
\usepackage{xcolor}
\usepackage{multirow}
\usepackage{booktabs}  
\usepackage{lipsum}
\usepackage{pdflscape}
\usepackage{longtable}
\usepackage{booktabs}
\usepackage{geometry}

\title[A FAST dark galaxy candidates catalogue]{FAST and Dark: A catalogue of Dark Galaxy Candidates within 50 Mpc}

\author[M. Monaci et al.]{Marco Monaci,$^{1}$\thanks{E-mail: mmonaci@swin.edu.au}
Duncan A. Forbes$^{1}$,
Jonah S. Gannon$^{1}$,
Bärbel S. Koribalski$^{2,3}$,
\newauthor
Kenji Bekki$^{4}$,
Jean P. Brodie$^{1}$,
and Warrick J. Couch$^{1}$
\\
$^{1}$Centre for Astrophysics and Supercomputing, Swinburne University of Technology, John Street, Hawthorn VIC 3122, Australia\\
$^{2}$CSIRO Space and Astronomy, P.O. Box 76, Epping NSW 1710, Australia\\
$^{3}$School of Science, Western Sydney University, Locked Bag 1797, Penrith NSW 2751, Australia \\
$^{4}$International Centre for Radio Astronomy Research (ICRAR), M468, The University of Western Australia, 35 Stirling Hwy, Crawley, WA 6009, Australia
}

\date{Accepted XXX. Received YYY; in original form ZZZ}

\pubyear{\the\year{}}

\begin{document}
\label{firstpage}
\pagerange{\pageref{firstpage}--\pageref{lastpage}}
\maketitle

\begin{abstract}
Using the first data release of the Five-hundred-meter Aperture Spherical radio Telescope (FAST) All-Sky \ion{H}{i} survey (FASHI), we compile a catalogue of 70 dark galaxy candidates (DGCs) within 50 Mpc. We select DGCs without an identified optical counterpart at a limiting $g$-band magnitude of $\sim$ 28 mag arcsec$^{-2}$ in the DESI Legacy Survey, using both automatic cross-checking with optical catalogues and visual inspection of the colour images. After validating our DGCs, excluding potential spurious detections, issues in the registered position of the \ion{H}{i} sources, and possible Radio Frequency Interferences (RFIs), we analyse their distribution over the surveyed sky, \ion{H}{i} mass, linewidths, and inferred distance. They appear evenly distributed across the surveyed area, with no apparent bias to isolation. We did not find any DGC within the Local Volume (11 Mpc) in the sky surveyed by this first release of FASHI. We compare the observed properties of DGCs with those of galaxies with optical counterparts, finding that DGCs tend to have higher linewidths for a given \ion{H}{i} mass. We discuss our DGCs in light of theoretical works, and compare them with other observational samples from previous \ion{H}{i} surveys. This work presents a catalogue of dark galaxy candidates, which can serve as a basis for follow-up studies.
\end{abstract}



\section{Introduction}
Although the $\Lambda$CDM model \citep[][]{1978MNRAS.183..341W, 1991ApJ...379...52W} successfully describes the hierarchical large-scale structure of the Universe, with haloes of dark matter (DM) spanning several orders of magnitude in mass, it also predicts a much larger number of low-mass haloes than are actually observed. This tension is, in general, explained by the fact that below a certain threshold mass haloes have their galaxy formation suppressed \citep[][]{1986MNRAS.218P..25R, 1996MNRAS.278L..49Q, 1999ApJ...523...54B, 2002MNRAS.336..541V, 2017MNRAS.465.3913B}, and the threshold at which this occurs remains an outstanding question in galaxy formation.

The reasons why a halo may fail to form stars efficiently have been extensively studied, considering various physical processes. Past works in general agree that there is a gas surface density threshold to ignite the star formation \citep[][]{1980MNRAS.193..189F, 1989ApJ...344..685K}, and that the ultraviolet (UV) background radiation may play a crucial role \citep[][]{1992MNRAS.256P..43E,2004ApJ...609..667S}.
Several theoretical studies \citep[i.e.][]{2008MNRAS.390..920O, 2020MNRAS.498.4887B, 2021ApJ...923...35M, 2025arXiv250609122D} indicate that only haloes with a mass above a certain threshold can keep the gas within their potential well and form stars. This critical halo mass is thought to be, at present day, $\sim 5 \times 10^9$ M$_{\odot}$; however, it has been shown that it depends on redshift and the epoch of reionisation. In fact, reionisation can suppress star formation by heating baryons and preventing their condensation to form stars \citep[][]{2000ApJ...539..517B, 2017MNRAS.465.3913B}.

Haloes that failed to form stars, or with very low star formation efficiency, are usually referred to as `dark galaxies', although a common definition is still lacking. Dark haloes with no gas or stars are believed to be common in the low-mass end of the distribution, but they are virtually undetectable since they do not host baryons. 
On the other hand, \citet{1997MNRAS.292L...5J} showed that several DM haloes host a rotating gaseous disc that cannot form stars because of their low surface density. However, they should be detectable through \ion{H}{i} observations since they retain a fair amount of neutral hydrogen \citep[][]{2006MNRAS.368.1479D, 2012AJ....144..159M,2015AJ....149...72C, 2020A&A...642L..10B}.

In recent years, a significant amount of work has been devoted to simulating these objects to understand their evolution and properties, as well as establishing observational constraints for their detection. \citet{2006MNRAS.368.1479D} simulated galactic discs embedded in Navarro, Frenk, and White \citep[NFW,][]{1996ApJ...462..563N} DM haloes, finding that these objects are expected to have \ion{H}{i} column densities between $0.5 - 5 \times 10^{20}$ cm$^{-2}$, sufficiently high to self-shield from photoionisation but sufficiently low not to form stars efficiently. The simulated dark galaxies have $6 < \log (M_{\rm HI}/M_{\odot})<~9$ and a velocity widths $\lesssim 100$ km s$^{-1}$, although the majority of them have \ion{H}{i} masses below 10$^{7}$ M$_{\odot}$ and velocity widths less than 40 km~s$^{-1}$.

\citet{2017MNRAS.465.3913B} used the \texttt{APOSTLE} simulation to examine the physical properties of low-mass haloes, finding a population free of stars with the gas in thermal equilibrium with the ionising UV background and in hydrostatic equilibrium with the dark matter gravitational potential. These low-mass dark galaxies are referred to by the authors as RELHICs (REionization-Limited \ion{H}{i} Clouds). RELHICs have the gas nearly fully ionised, rendering them difficult to observe; however, a central core may be composed of neutral hydrogen, so the massive tail of the distribution ($4.5 \lesssim \log(M_{\rm HI}/M_{\odot})\lesssim 6.5$) should be detectable by extragalactic \ion{H}{i} blind surveys.

\citet{2020MNRAS.498L..93J}, using their previous model \citep[][]{1997MNRAS.292L...5J}, suggest that a population of high-spin dark galaxies could exist. These systems would be Toomre-stable and are expected to have larger sizes than luminous galaxies at fixed halo mass, preferentially residing in isolated environments.

More recently, \citet{2024ApJ...962..129L} used the Illustris TNG50 \citep[][]{2019MNRAS.490.3234N} cosmological hydrodynamical simulation to study the formation and evolution of dark galaxies. They predicted that dark galaxies are characterised by larger sizes, higher halo spin parameters, tend to reside in isolation, and have experienced fewer mergers throughout their formation history compared to luminous galaxies. However, they do not explicitly address the \ion{H}{i} content of these systems, focusing instead on the total gas mass (warm, cold and star-forming), which makes the comparison with observations difficult.

Despite theoretical work carried out in the past, definitive observational confirmation of dark galaxies remains lacking. However, the most effective strategy appears to be searching for 21 cm radio emission from neutral hydrogen over broad areas of the sky.

The \ion{H}{i} Parkes All Sky Survey \citep[HIPASS,][]{2004MNRAS.350.1195M, 2006MNRAS.371.1855W} has covered all the southern sky up to a declination of +25\degr, with a spatial (regridded) resolution of 15\farcm5, a sensitivity of 13 mJy beam$^{-1}$, and was expected to reveal several starless haloes, but no convincing candidates were found.  \citet{2005MNRAS.361...34D} carefully analysed the HIPASS catalogues, finding only 13 promising candidates, 12 of them excluded by subsequent follow-up, with the remaining one having a faint optical counterpart. 

The Arecibo Legacy Fast ALFA survey \citep[ALFALFA,][]{2005AJ....130.2598G, 2018ApJ...861...49H, 2011AJ....142..170H} has covered 7000 deg$^{2}$ on the northern sky, with a spatial resolution of 3\farcm3 $\times$ 3\farcm8, and with a better sensitivity than HIPASS ($\sim 2$ mJy beam$^{-1}$). \citet{2025ApJS..279...38K} recently analysed the ALFALFA survey, claiming 142 dark galaxy candidates without an evident optical counterpart in the DESI Legacy Imaging Surveys \citep[hereafter, Legacy;][]{2019AJ....157..168D, 2022AJ....164..207D}.

Single-dish surveys are limited by their spatial resolution. This issue is overcome by the Widefield ASKAP L-band All-Sky Blind surveY \citep[WALLABY,][]{2020Ap&SS.365..118K} that takes advantage of the Australian SKA Pathfinder's (ASKAP's) ability to survey large areas of sky at high spatial (30\arcsec) resolution and with a sensitivity of 1.6 mJy beam$^{-1}$. \citet{OBierne2023}, and recently revised by \citet{2025PASA...42...87O}, examined the pilot fields of WALLABY, claiming two dozen dark galaxy candidates, although they require deep follow-up \ion{H}{i} observations to definitively confirm the detections.

In addition to being intrinsically difficult to observe, dark galaxy candidates should also be relatively isolated; observationally, this may be required to avoid \ion{H}{i} tidal tails. \citet{2020arXiv200207312K} discussed and reviewed \ion{H}{i} structures in interacting pairs of galaxies, which show very low surface brightness optical counterparts, or even none. These structures could extend over 150 kpc away from the interacting pair, and, in principle, could mimic the \ion{H}{i} emission of a dark galaxy. 

An emblematic example is the source VIRGOHI21, first claimed to be a dark galaxy \citep[][]{2004MNRAS.349..922D, 2005ApJ...622L..21M}, and after further study was reconsidered to be interaction-based tail \citep[][]{2005MNRAS.363L..21B, 2007ApJ...665L..19H, 2008ApJ...673..787D}.

\citet{2023ApJ...952..130Z} performed high-sensitivity \ion{H}{i} observations toward M94 with the Five-hundred-meter Aperture Spherical radio Telescope \citep[FAST,][]{2011IJMPD..20..989N, 2019SCPMA..6259502J}, detecting several \ion{H}{i} clouds, one of which (Cloud-9) lies 109 kpc from M94. \citet{2024ApJ...973...61B} subsequently carried out interferometric observations of Cloud-9 with the Very Large Array (VLA), showing that it is non-rotating and dynamically cold. Follow-up HST observations by \citet{2025ApJ...993L..55A} ruled out the presence of optical counterparts down to stellar mass $\sim 10^4$ M$_{\odot}$. However, given its projected distance from M94 and its similar radial velocity, Cloud-9 is not an optimal candidate for an isolated dark galaxy.

Deep optical surveys are crucial to search for potential optical counterparts. In fact, previously identified good dark galaxy candidates can turn out to be extremely low-surface-brightness galaxies, as revealed by deeper surveys. For example, the \ion{H}{i} source AGC 229101, initially classified as a dark galaxy candidate, turns out to have a faint optical counterpart in a deep image ($\sim 28.5$ mag arcsec$^{-2}$, estimated stellar mass of $\sim 10^7$ M$_{\odot}$) by the WIYN telescope at Kitt Peak National Observatory \citep[][]{2021AJ....162..274L}.
Similarly, \citet{2024ApJ...964...85D} identified in the Legacy Survey eight optical counterparts of \ion{H}{i} sources detected in ALFALFA but with no visible counterparts in Sloan Digital Sky Survey (SDSS) images, which has a shallower depth \citep[23–23.5 mag arcsec$^{-2}$,][]{2005ApJ...631..208B}.

Motivated by the need for more high-quality dark galaxy candidates, in this work, we will use the data from FAST. FAST will observe the entire sky in \ion{H}{i} between -14\degr ~and +66\degr ~in declination, and create an unprecedented blind extragalactic \ion{H}{i} survey. Given the telescope's large aperture, the survey will have excellent angular resolution (3\arcmin, although still lower than that of interferometric surveys) and with a sensitivity of 0.76 mJy beam$^{-1}$ at a spectral resolution of 6.4 km s$^{-1}$, it will be deeper than previous surveys, and is expected to detect hundreds of dark galaxy candidates. 

\citet{2024SCPMA..6719511Z} presented the first release of the FAST All Sky \ion{H}{i} (FASHI), which is the catalogue we use in this study. 
The FAST telescope has already shown its potential to detect faint \ion{H}{i} sources: \citet{2024A&A...684L..24K}, using the FASHI catalogue, discovered 20 nearby \ion{H}{i} sources linked with faint dwarf galaxies, together with seven dark clouds with no optical counterpart, in nearby groups of galaxies. \citet{2025ApJS..278...37Y} found seven \ion{H}{i} sources without an optical counterpart in the Ursa Major Supergroup, \citet{2022RAA....22f5019K} analysed the first pilot extragalactic survey made by FAST, finding 16 \ion{H}{i} sources without any optical counterpart, but only three are within 50 Mpc and Legacy images are not available for them. 
\citet{2023ApJ...944L..40X} identified an isolated FASHI source with apparently no optical counterpart on relatively shallow Pan-STARRS1 images (surface brightness limit of 23-23.5 mag arcsec$^{-2}$), and claimed it to be a good dark galaxy candidate.
Recently, \citet{2026A&A...708A..40S}, using the VLA, obtained a refined \ion{H}{i} position, leading to the discovery of a faint optical counterpart in Pan-STARRS1 images that was not identified by \citet{2023ApJ...944L..40X}, likely due to the lack of a precise position given the large FAST beam. Concurrently, \citet{2026A&A...705L...9M} obtained deep optical imaging and spectroscopic follow-up, unambiguously confirming the presence of an optical counterpart.

For the purposes of our work, we define a Dark Galaxy Candidate (DGC) as an isolated extragalactic \ion{H}{i} detection (at least 150 kpc away from any bright galaxy) with no visible optical counterpart in the Legacy {\it g}-band down to $\sim$28 mag arcsec$^{-2}$. The paper is organised as follows. In Section \ref{sec:catalogues}, we describe the catalogues we used to filter out the known luminous galaxies. In Section \ref{sec:selection_criteria} we describe the selection criteria and the workflow we applied. In Section \ref{sec:validation} we discuss and validate the DGCs final catalogue. In Section \ref{sec:discussion} we discuss the properties of the DGCs. 
We draw our conclusions in Section \ref{sec:conclusions}.

\section{Catalogues and surveys}\label{sec:catalogues}
In this Section, we will briefly discuss the catalogues used for source selections.

\subsection{The FAST All Sky HI Survey (FASHI)}\label{subsec:fashiCat}
FAST is, to date, the largest and most sensitive single-dish radio telescope. Its frequency range spans from 70 MHz to 3000 MHz. It has an effective aperture of 300 meters and can observe and track any sources within 40 degrees from the zenith. Alongside other projects, FAST will observe the entire sky visible from the telescope location in the frequency range of 1000 -- 1500 MHz. It will observe both Galactic and extragalactic \ion{H}{i} sources with a velocity resolution of 6.4 km s$^{-1}$, a median detection sensitivity of 0.76 mJy beam$^{-1}$ at 6.4 km s$^{-1}$ resolution, and a beamsize of $\sim$3\arcmin \ diameter \citep[][]{2020RAA....20...64J}.

The first data release of the FAST All Sky \ion{H}{i} survey \citep[FASHI,][]{2024SCPMA..6719511Z} primarily focuses on extragalactic sources outside the region of the sky observable by the Arecibo telescope. It covers two main areas of the sky, one between +30$\degr$ and +66$\degr$ in declination, and another thin strip around the celestial equator between -6$\degr$ and 0$\degr$. In right ascension, the covered areas are between $0^{\rm h}-17.3^{\rm h}$ and $22^{\rm h} - 24^{\rm h}$. The Galactic plane is partially covered between $\ell = 130\degr$ and $\ell = 180\degr$. The catalogue consists of 41741 \ion{H}{i} sources detected between 1305.5 -- 1419.5 MHz. Radio Frequency Interference (RFI) mitigation involved several strategies, including avoiding polluted bands and limiting observations to frequencies above 1305 MHz, which are relatively clear. Standing waves, i.e., baseline ripples produced by internal reflections in the telescope optics, are also mitigated.

The source extraction workflow involves both automatic and interactive procedures. After a first source selection of potential candidates over 4.5$\sigma$, a manual validation procedure was performed, flagging and removing as much as possible occasional RFIs, OH megamasers, and radio recombination lines \citep[for further details, see Sections 3 and 4 of][]{2024SCPMA..6719511Z}. The final catalogue comprises only the sources with a spectral S/N $\geq$ 5. Approximately 2000 sources with a lower S/N were excluded.
The FASHI catalogue lists, for each \ion{H}{i} source, the position (RA and Dec), the heliocentric velocity (which we indicate, hereafter, as $cz_{\odot, \rm radio}$), distances (inferred from a cosmic flow model) and properties linked with the shape of the line, as the linewidth at 50\% of the peak flux W$_{50}$, the linewidth at 20\% of the peak flux W$_{20}$, the peak flux density F$_{\rm P}$, and the S/N.

\subsection{The 50 Mpc catalogue (50MGC)}
The 50MGC catalogue \citep[][]{2024AJ....167...31O} lists 15424 nearby galaxies within 50 Mpc. The catalogue combines data from various other catalogues, such as the NASA-Sloan Atlas (NSA)\footnote{\url{https://nsatlas.org/}}, HyperLEDA\footnote{\url{http://leda.univ-lyon1.fr/}} \citep[][]{2014A&A...570A..13M}, and the Local Volume Galaxies catalogue \citep[][]{2013AJ....145..101K}. 
It is worth mentioning here that NSA is based on the Sloan Digital Sky Survey (SDSS), which has a surface brightness limit of about 23-23.5 mag arcsec$^{-2}$ \citep[][]{2005ApJ...631..208B}.

\subsection{The Siena Galaxy Atlas (SGA)}
The Siena Galaxy Atlas \citep[SGA, see][]{2023ApJS..269....3M} was created using the optical \textit{grz} images from Legacy and the four infrared bands (3.4, 4.6, 12, and 22 $\mu$m) from the WISE survey \citep[][]{2010AJ....140.1868W} and the WISE-Reactivation Mission \citep[][]{2014ApJ...792...30M}. The atlas contains 383620 nearby galaxies, and for each of them lists the position, ellipticity, surface brightness, as well as a cross-match with the HyperLEDA catalogue and optical redshifts.

\subsection{The Systematically Measuring the Ultra-Diffuse Galaxies catalogue (SMUDGes)}
The Systematically Measuring Ultra-diffuse Galaxies catalogue \citep[SMUDGes, see][]{2023ApJS..267...27Z} is devoted to specifically listing low surface brightness galaxies, and therefore is useful to exclude faint sources from our selection. The creation of the SMUDGes catalogue is fairly complex and involves several steps, including an automatic detection pipeline performed over the {\it g}, {\it r}, and {\it z} bands of the DR9 northern portion of the Legacy.
The final catalogue contains 7070 Ultra Diffuse Galaxies (UDGs) candidates, of which 1529 have distance estimations.

\subsection{The NASA/IPAC Extragalactic Database (NED)}\label{subsec:NED}
The NASA/IPAC Extragalactic Database (NED\footnote{\url{https://ned.ipac.caltech.edu/}}) is an online database that collects and cross-matches nomenclature, radial velocities, distances and several other parameters of hundreds of millions of extragalactic sources. This resource should not be regarded as a homogeneous catalogue, like the ones described in the previous sections, as the data come from different surveys and telescopes.

\subsection{The DESI Legacy Imaging Surveys}\label{subsec:DESI}
The DESI Legacy Imaging Surveys \citep[][]{2019AJ....157..168D} is the combination of three different optical surveys, namely the Dark Energy Camera Legacy (DECaLS), the Beijing-Arizona Sky Survey (BASS), and the Mayall z-band Legacy (MzLS). Together, the three surveys cover an area of 14000 deg$^2$ of the northern hemisphere, without covering the Galactic plane. The DESI project aims to have a depth as uniform as possible across the sky surveyed. MzLS and BASS observed regions with a declination $\geq 32\degr$, while DECaLS observed regions below 32$\degr$. The DECaLS survey is about 0.5 mag deeper than BASS+MzLS. \citet{2023A&A...671A.141M} indicate $\sim 28.4$ mag arcsec$^{-2}$ as the average surface brightness limit calculated over an area of 10\arcsec $\times$ 10\arcsec at 3$\sigma$ in the DESI Legacy Imaging Surveys (see their figure 12). In this work, we used the 10$^{\rm th}$ Data Release (DR10), when available; otherwise, we used the DR9.

\subsection{Radio catalogues: the Arecibo Legacy Fast ALFA survey (ALFALFA)}\label{subsec:radiocatALFALFA}
The Arecibo Legacy Fast ALFA survey \citep[ALFALFA, for a full discussion see][]{2005AJ....130.2598G, 2005AJ....130.2613G,2011AJ....142..170H,2018ApJ...861...49H} was conducted using the W. Gordon Telescope (Arecibo), surveying a substantial fraction (about 7000 deg$^2$) of the sky visible from Arecibo, avoiding the Galactic equator. The velocity resolution is $\sim 10$ km s$^{-1}$ for a bandwidth between 1335 MHz and 1435 MHz (highest $z = 0.06$), and the beamsize at 1420 MHz is about 3\farcm3 $\times$ 3\farcm8, which is compatible with the angular resolution of the FAST telescope ($\sim$ 3\arcmin). The catalogue contains about 31500 extragalactic sources with a redshift $z_{\odot} < 0.06$ but only a relatively small stripe between Dec = +30\degr and Dec = +36\degr overlaps with the FASHI coverage. 

\subsection{Radio catalogues: the \ion{H}{i} Parkes All Sky Survey (HIPASS)}\label{subsec:radiocatHIPASS}
The \ion{H}{i} Parkes All Sky Survey \citep[HIPASS, see][]{2004MNRAS.350.1195M, 2004MNRAS.350.1210Z, 2004AJ....128...16K, 2005MNRAS.361...34D, 2006MNRAS.371.1855W} was conducted using the Parkes Telescope and covers the whole sky visible from the telescope location. The survey is divided into two sub-surveys: the southern survey covers the sky between +2\degr and -90\degr in declination, while the northern survey covers the sky between +25\degr and +2\degr in declination. The velocity range is from -1280 to 12700 km s$^{-1}$, which coincides with a maximum redshift of z $\sim 0.043$, and with a velocity resolution of 18 km s$^{-1}$. The average beam size (regridded) is 15.5\arcmin, significantly larger than the FAST beam. The final HIPASS and northern HIPASS catalogues comprise, respectively, 4315 and 1002 high-fidelity extragalactic sources. Only the equatorial band observed by FAST, from –6\degr to 0\degr in declination, is shared between the HIPASS and FASHI surveys.

\section{Selection criteria}\label{sec:selection_criteria}
\begin{figure}
    \includegraphics[width=\columnwidth]{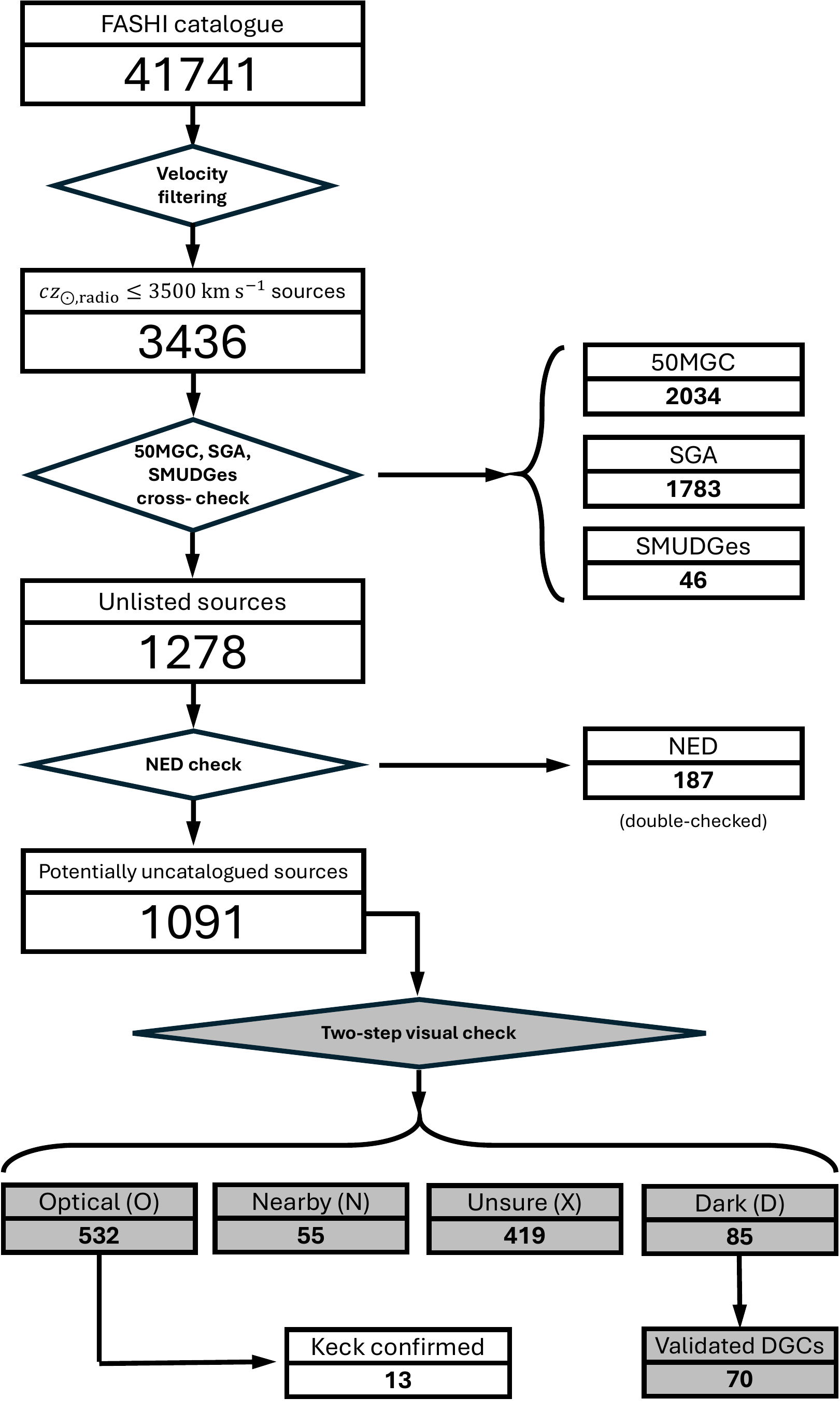}
    \caption{Flowchart describing the applied filtering criteria. All the steps with the white background were performed using objective criteria, as described in the text. Therefore, these numbers can be easily reproduced by applying the same procedure. We double-checked the 187 NED sources to see if we excluded some valid DGCs, but none of them meet the conditions we imposed to be a reliable DGC (see Section \ref{subsec:ned_check}). The steps with a grey-shaded background were performed by visually inspecting each galaxy of the sample through the Legacy, and after the validation procedure. For a comprehensive discussion of the classes into which we divided the sources based on the visual check, see Section \ref{subsec:visual_check}.}
    \label{fig:flowchart}
\end{figure}

In this Section, we describe our selection process, tracing the process from the FASHI catalogue through to the final list of DGCs. Figure \ref{fig:flowchart} shows the steps we followed for the selection.

\subsection{Velocity filtering}\label{subsec:velocity_filtering}
The full FASHI catalogue includes sources with velocities ranging from $\sim$ 200 km s$^{-1}$ to $\sim$ 26323 km s$^{-1}$. We limited our search for nearby galaxies to within 50 Mpc. Adopting H$_{0} = 70$ km s$^{-1}$ Mpc$^{-1}$, this yields a velocity limit of $\sim 3500$ km s$^{-1}$. Therefore, we first applied a velocity filter of $cz_{\odot, \mathrm{radio}} \leq 3500$ km s$^{-1}$ to the FASHI catalogue, obtaining a filtered catalogue of 3436 \ion{H}{i} sources (see Figure \ref{fig:flowchart}, first step). It's worth mentioning here that this cut in velocity, in principle, protects our filtered catalogue from the majority of RFI, since the frequency range of 1403 MHz to 1419.5 MHz falls within the protected band for radio astronomy, where the 21 cm line may reside.

\subsection{Cross-matching with optical catalogues}\label{subsec:matching_optical}
After obtaining the velocity-filtered catalogue, we performed a cross-matching with the 50MGC, SGA, and SMUDGes catalogues in RA/Dec. For single-dish surveys, like the FAST telescope, the centroiding accuracy can be estimated by dividing the beam size (3\arcmin \ for FAST) by the S/N of the detection \citep[][]{2004AJ....128...16K}. Since the FASHI catalogue has an S/N threshold of 5, the worst-case centroiding accuracy is 0\farcm6 in radius, considering a circular beam. However, as pointed out by \citet{2024SCPMA..6719511Z}, analysing the offset between the \ion{H}{i} detections and their optical counterparts in the Siena Galaxy Atlas, the centroiding accuracy could be worse than expected even for the brighter sources, on the order of 1\arcmin or even more. For example, one of our spectroscopically confirmed optical counterparts (see Section \ref{app:KeckSpectra}, source J080428.28+405843.1) is located about 1\farcm3 \ from the \ion{H}{i} listed position. We decided to maintain a conservative approach: cross-matching is based only on RA/Dec positions, searching for galaxies within 1\farcm5 in radius, without imposing any redshift constraint, since redshift information is not available for all the galaxies listed in the optical catalogues. This approach may exclude sources without a true physical association, but it avoids including those that actually are physically related.

After this cross-matching, of the 3436 \ion{H}{i} sources, 2158 were found to have an optical counterpart. In particular, 2034 were found in the 50MGC catalogue, 1783 in the SGA catalogue, and 46 in the SMUDGes catalogue. It is worth noting that these numbers don't add up to 2158, since several sources were found in more than one catalogue (see the upper part of the flowchart in Figure \ref{fig:flowchart}). Therefore, 1278 unique FASHI sources were not found in any of the aforementioned catalogues.

\subsection{Cross-matching with the NED database}\label{subsec:ned_check}
To further improve our selection, we performed a search in the NED database for each source. Given the considerable number of listed objects in the NED database and the relatively large beam size of the FAST radio telescope, performing a search based solely on position may produce misleading results, identifying several background sources that are not associated with the \ion{H}{i} source. To mitigate this problem, we queried the NED database, adding a condition on the radial velocity of the potential optical counterpart. To be selected as a counterpart, a galaxy must lie within 1\farcm5 ~radius of the queried position and must have a velocity within $cz_{\odot, \mathrm{radio}} \pm 200$ km s$^{-1}$ of the radio detection. If the velocity is not available for any galaxy within the search radius of 1\farcm5, then no galaxy is selected as the optical counterpart. Although this stringent condition does not allow for definitive identification of whether a source has been catalogued, as the correct optical counterpart might lack a radial velocity measurement, it serves to minimise the risk of spurious associations (an example is described in Appendix \ref{app:optCountQuest}).

After our check, out of 1278 sources, we found 187 in the NED database: most of them are catalogued in SDSS (Sloan Digital Sky Survey), WISEA (Wide-field Infrared Survey Explorer All-sky), HIZOA (\ion{H}{i} Zone of Avoidance), EZOA (Effelsberg survey of \ion{H}{i} Zone of Avoidance), and GALEXASC (GALEX All-sky Survey Catalogue). This step may indeed have excluded some valid DGCs previously identified in radio catalogues. We therefore visually checked the \ion{H}{i} sources, but none of them meets the conditions we describe in the next Section to be classified as a reliable DGC. In this step, we excluded all new sources from \citet{2024A&A...684L..24K}, i.e. the added \ion{H}{i} sources to the Local Volume, as they either have an optical counterpart or lie within 150 kpc of a bright galaxy.
The remaining 1091 sources were visually inspected, as described in the next subsection.

\subsection{Visual check of uncatalogued galaxies}\label{subsec:visual_check}
After cross-checking our source list with optical catalogues, we obtained a sample of 1091 objects potentially not previously catalogued. However, some of these sources may appear in catalogues not included in our analysis, as no complete and homogeneous catalogue of all known extragalactic sources currently exists. Our aim is to have a visual assessment that is easily accessible to others for future studies and follow-up, identifying candidates that are suitable for deeper optical or UV imaging or other detection techniques. 

The visual inspection of the sources has been performed using a two-step approach. For the initial screening, we created a Python script that, for a given position, displays a full-colour optical Legacy image (from DR10, where available; otherwise, from DR9), with dimensions of 5\arcmin $\times$ 5\arcmin. The size of the cutout was deliberately chosen to be wider than the FAST beam, to be sure not to miss any potential optical counterpart that could be poorly spatially matched with the \ion{H}{i}-inferred position. The first step was intended as a preliminary selection to discard clear optical counterparts. We note that BASS+MzLS and DECaLS (see Section \ref{subsec:DESI}) have different surface-brightness sensitivities, and they change even within the same survey. It is also important to note that the colour information was very useful for identifying galaxies over stars and other non-galactic features in the images, whereas the grey images alone are not as informative.

For each entry, the script allows for classification of the source, selecting one of the following categories:

\begin{itemize}
    \item Optical counterpart (O): the optical Legacy image is available and shows a potential optical counterpart within 1\farcm5 ~from the \ion{H}{i} central position. A typical optical counterpart for the FASHI sources is a small blue galaxy with few or no features, very near the centre of the beam. We nonetheless emphasise here that this classification is inherently subjective, and it is possible that some sources are not physically associated with the \ion{H}{i} emission (see the case discussed in Appendix \ref{app:KeckSpectra}). However, in the majority of cases, the optical counterparts appear rather unambiguous;
    
    \item Nearby (N): if there is no clear optical counterpart in the image, but there is a luminous and large galaxy within 150 kpc in radius at the distance of the FASHI source and sharing a radial velocity compatible with that of the FASHI source (within $\pm$200 km s$^{-1}$), we visually classify the source in this category. In these cases, the \ion{H}{i} source could be a tidal tail of the nearby galaxy \citep[][]{2020arXiv200207312K}. We note here that for every galaxy in our sample, the 1\farcm5  ~radius beam is always smaller than the fiducial radius of 150 kpc; the distance at which the beam encloses exactly this diameter is about 344 Mpc, well beyond our limit of 50 Mpc. We should note here that some sources in this category could be perfectly legitimate dark galaxy candidates, but since we want to get a catalogue that is as reliable as possible, we remove such sources;
    
    \item Unsure (X): We flagged as "unsure" all sources for which the optical image from Legacy is unavailable or for which we are uncertain, i.e., there was not a clear optical counterpart near the centre, or the potential optical galaxies that could be the optical counterparts were quite off the centre of the beam. In these cases, the examined source falls into this category;
    
    \item Dark (D): In this category fall all the \ion{H}{i} sources without a clear optical counterpart, and well isolated without known bright galaxies within 150 kpc at the distance of the FASHI source and with a similar recessional velocity (within $\pm$200 km s$^{-1}$) as the \ion{H}{i} detection.
\end{itemize}

\begin{figure*}
    \includegraphics[width=\textwidth]{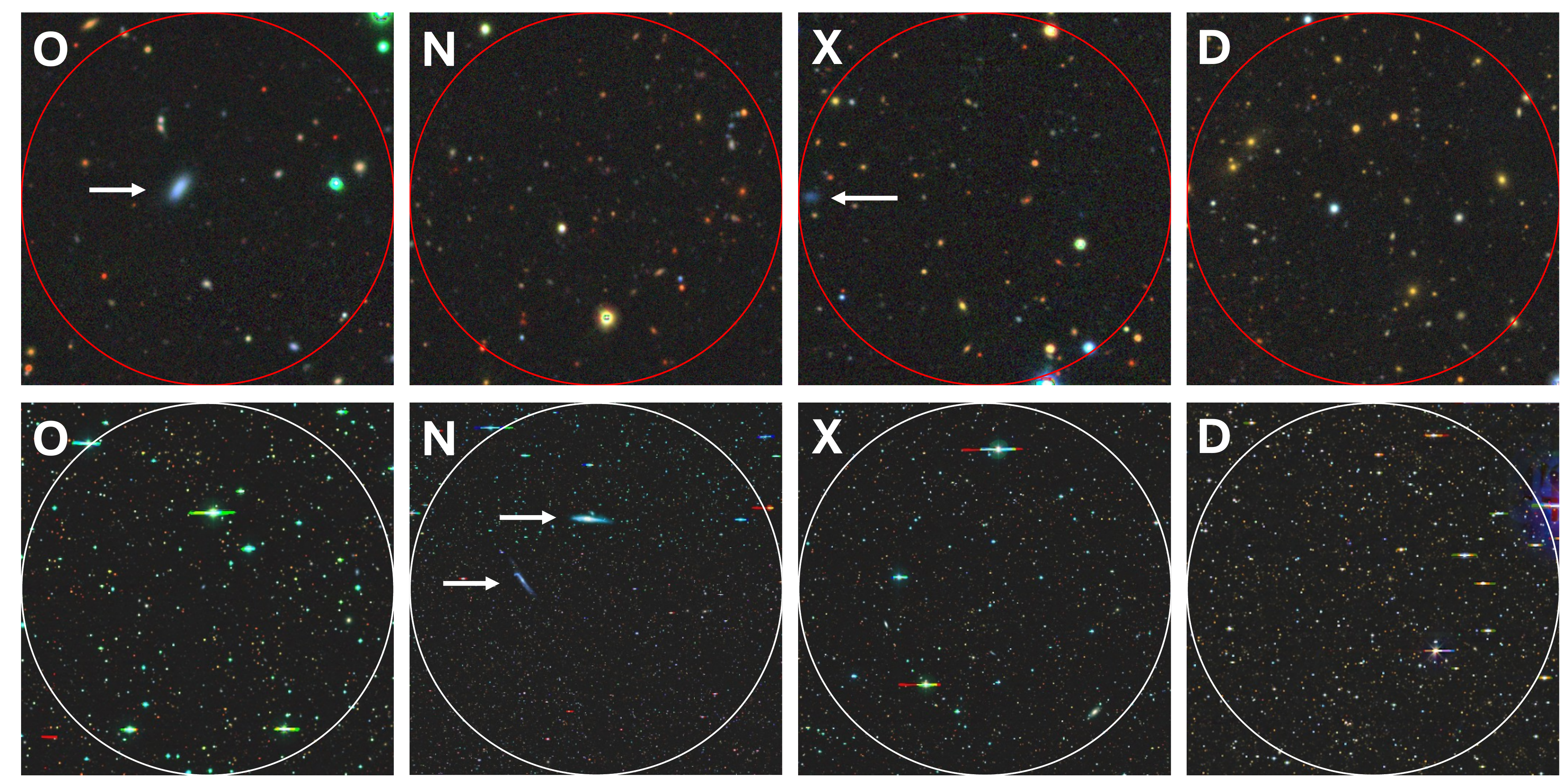}
    \caption{From left to right, examples of the four categories: optical counterpart (O), nearby galaxy (N), unsure sources (X), dark galaxy candidates (D). The upper row shows cutouts with the size of the FAST beam (3\arcmin, identified by the red circle), and the lower row shows cutouts with a 150 kpc-radius circle (in white). Since the 150 kpc-radius circle vary in size given the different distances of the FASHI sources, the field of view of the lower cutouts differs from each other.
    In all images, the FASHI source is located in the centre of the cutout. \textit{(O - Optical Counterpart.)} The blue galaxy, identified by a white arrow, represents the optical counterpart of the FASHI source J165132.43+581535.3, also confirmed by a radial velocity measurement obtained using KCWI at Keck (see Appendix \ref{app:KeckSpectra}). \textit{(N - Nearby Galaxy.)} The FASHI source is J124154.31+320735.2, which is located near NGC 4656 (the galaxy on the left in the 150 kpc cutout highlighted by a white arrow) and NGC 4631 (on the top). The FASHI source has $cz_{\rm \odot, radio}\simeq 670.83$ km s$^{-1}$, while NGC 4656 has $cz_{\odot}\simeq 646$ km s$^{-1}$, and NGC 4631 has $cz_{\odot} \simeq 616$ km s$^{-1}$. Given the vicinity and the very similar radial velocities, the \ion{H}{i} source may be a tidal debris or a gas tail. 
    \textit{(X - Unsure Source.)} The FASHI source J160441.93+414207.2 does not present a clear optical counterpart in the 3\arcmin ~cutout, other than the small galaxy on the left highlighted by a white arrow, for which an optical redshift is not available. Given the lack of information about the radial velocity and the poor position matching, we classified this source as unsure.
    \textit{(D - Dark Galaxy Candidate.)} A good dark galaxy candidate (FASHI source J033817.46+001508.0); both cutouts show no clear optical counterpart, except for some background galaxies.}
    \label{fig:exampleClass}
\end{figure*}

Figure \ref{fig:exampleClass} shows an example for each category. 
After this first screening, we proceeded to inspect in more detail those sources with no evident optical counterpart. To do this, we downloaded the $g$, $r$, $i$ images from the DESI Legacy, stacking and smoothing them to enhance the visualisation of potential low-brightness counterparts. 
We attempted to smooth the images using a 2D Gaussian kernel with different $\sigma$ values, namely 3\arcsec, 5\arcsec, and 7\arcsec (note that 1 kpc in size is $\sim$7\arcsec at 30 Mpc and $\sim$4\arcsec at 50 Mpc). However, at these smoothing scales, sources were smeared and merged with the background noise, making optical assessment extremely difficult without the additional information provided by the full-colour images. Therefore, we decided to inspect the unsmoothed images. Each stacked cutout has been checked using different dynamic scales to enhance the potential faint optical counterpart. No source classified as lacking an optical counterpart in the full-colour images (first step) revealed a faint counterpart in the stacked grey-scale images.

After this double-step procedure, we identified 532 sources with a clear optical counterpart, 55 that show one or more luminous galaxies within 150 kpc, 419 that were classified as unsure sources, and 85 that remained as DGCs.

However, we emphasise that the search for optical counterparts of \ion{H}{i} sources detected through single-dish surveys is a long-standing problem in the literature, given the poor positional precision on the sky (3\arcmin \ for the FAST beam), which usually encompasses several possible optical counterparts. Without redshift information, only interferometric \ion{H}{i} observations could pinpoint a potential optical counterpart within $\sim$15\arcsec -- 30\arcsec. Although we performed the visual inspection with the greatest possible care, it is possible that some optical counterparts are present in the Legacy imaging but could not be unambiguously identified, and therefore, further observations are required (better \ion{H}{i} centroiding, deeper optical observations).

\subsection{Cross-matching with radio catalogues}\label{subsec:matching_radio}
We also performed a cross-match with the ALFALFA and HIPASS catalogues on the velocity-filtered \ion{H}{i} source list, simultaneously with the cross-matching with the optical catalogues. If a FASHI source is also found in one or both catalogues, it is simply flagged but not excluded from the catalogue.

The cross-matching with ALFALFA and HIPASS follows the same approach as we applied to the optical catalogues, but using different search radii, since the positions of the ALFALFA sources, as well as the HIPASS sources, are not known with the same precision as optical sources. In the case of ALFALFA, since the beamsize is reasonably similar to that of the FAST telescope (3\farcm3 $\times$ 3\farcm8 ~compared with 3\arcmin), we used a search radius of 3\arcmin \ (so twice the radius of the single beams). The case of HIPASS is different, since the survey has a regridded beam size of $\sim$ 15\farcm5. In this case, we used half the beam size (7\farcm25) as the search radius. We also tried a 9\arcmin \ radius search (7\farcm5 + 1\farcm5), but it led to spurious associations. The cross-matching is solely based on position, without any constraint on redshift. To maintain a conservative approach, we decided to perform the cross-matching with the radio catalogues using the full beam radius (see the discussion in Section \ref{subsec:visual_check}).
After this procedure, 406 FASHI sources were found in the ALFALFA catalogue and 189 in the HIPASS catalogue.
This cross-matching is not included in Figure \ref{fig:flowchart}, since it was not an applied filter. None of our remaining DGCs is listed in ALFALFA, nor in the HIPASS catalogue; however, we checked the HIPASS spectra database for the 13 DGCs located within the HIPASS footprint for low S/N spectral features. In two cases, we found a very weak emission at the expected recessional velocities, which are not listed in the final HIPASS catalogue due to low S/N.

\subsection{Validation of some optical counterparts through optical spectroscopy }\label{subsec:HIvel_optvel}
The spectroscopic confirmation of 14 selected targets from sources with optical counterparts but which do not appear in any catalogue (class `O', final white step in Figure \ref{fig:flowchart}) was carried out with the Keck Cosmic Web Imager (KCWI) on the Keck II telescope. These were twilight `filler' targets over three nights. A complete description of the acquisition and data reduction process can be found in Appendix \ref{app:KeckSpectra}. We selected the most obvious optical counterpart to the HI source, i.e. generally a blue luminous galaxy near the beam centre.

For each galaxy, we fitted the spectrum using \texttt{pPXF} \citep[][]{Cappellari2023} to derive an optical radial velocity, which we then compared with the \ion{H}{i} velocity. If the two velocities were reasonably consistent (i.e. within 2$\sigma$), we could confidently identify the `luminous' galaxy as the optical counterpart of the \ion{H}{i} emission listed in the FASHI catalogue.

Of the 14 sources for which we obtained an optical radial velocity, 13 were found to be the FAST \ion{H}{i} source counterpart, as the optical and radio velocities agreed within $2\sigma$ (see Figure \ref{fig:velocity_offsets}). 
For the remaining source, however, the optical radial velocity is significantly higher, indicating that the galaxy initially identified as the potential optical counterpart is not associated with the \ion{H}{i} emission, but is instead a background galaxy (see Figure \ref{fig:J150625_case}). 
This case highlights that some luminous galaxies in the `O' class could be misidentifications. We further discuss in Appendices \ref{app:KeckSpectra} and \ref{app:optCountQuest} the selection of the optical counterparts.

\section{Validation of the final DGC catalogue}\label{sec:validation}
After cross-matching with several optical catalogues and double-checking the results with the NED database, we obtained a catalogue of 85 DGCs. However, upon analysing the properties of the sources, we identified a few issues in the FASHI catalogue, which led us to exclude 15 potentially spurious sources, leading to a final catalogue of 70 DGCs. In this section, we describe and discuss some suspicious cases, noting that others may still be present.

\subsection{Sources with incorrect listed position and velocity}\label{subsec:OffsetPos}
Five DGCs, located at Dec = -5\degr, appeared suspiciously close to bright galaxies, within just a few arcminutes and with an offset almost exclusively in right ascension. These \ion{H}{i} detections were initially considered valid, as the nearby bright galaxies had been excluded based on their significantly higher radial velocities (above our threshold of 3500 km s$^{-1}$), and thus the \ion{H}{i} detections could not be physically associated with them. However, the nearby bright galaxies were not actually detected by FAST, but rather by the HIPASS survey, or other previous \ion{H}{i} observations. After analysing the FAST spectra of these suspicious sources, we noted that the line shape is almost identical to what was reported by HIPASS. Therefore, these \ion{H}{i} detections turned out to be the \ion{H}{i} emission associated with the nearby galaxies, listed in the FASHI catalogue with a variable positional offset of a few arcmins (1\arcmin -- 5\arcmin) to the west, caused by a clock issue and with an incorrect velocity of about 2000 km s$^{-1}$ (C.P. Zhang, private communication). This problem is believed to affect only a limited number of sources at Dec $= -5$\degr ~that were probably observed during the same observational run and will be fixed in the next release.
Given these considerations, we decided to exclude these five sources at the same declination of -5\degr, with 4\degr < RA < 14\degr, that are located to the west of a bright galaxy detected by previous \ion{H}{i} surveys but {\it not} by FAST, and with a velocity offset compatible with 2000 km s$^{-1}$. In Appendix \ref{app:posIssues} we discuss these cases.

\subsection{The case of a grouping of DGCs}\label{subsec:dgccluster}
\begin{figure}
    \includegraphics[width=0.9\columnwidth]{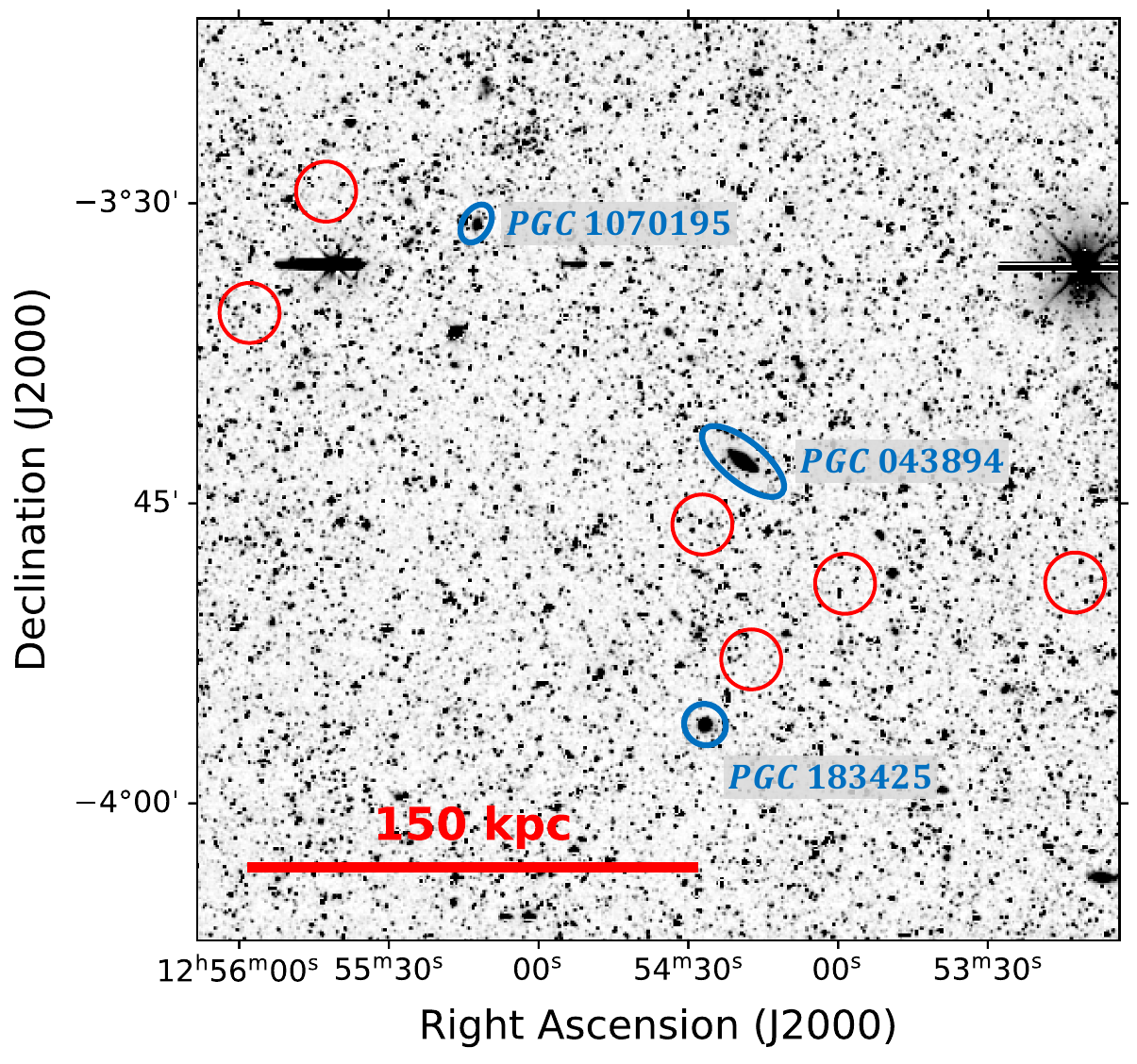}
    \caption{Cutout obtained by averaging the Legacy images in \textit{g,r,i,z} filters of the field around PGC 043894, PGC 1070195, and PGC 183425 (highlighted by blue ellipses). The field of view is about 0\fdg77 $\times$ 0\fdg77, and red circles (with the same diameter as the 3\arcmin \ beam width of the FAST telescope) highlight the six DGCs. 
    All the DGCs are located at almost the same distance of 23--24 Mpc, while the PGC galaxies are all in the background at distances $\geq$ 98 Mpc. FAST correctly identified PGC 043894. The 150 kpc bar assumes a distance of 23 Mpc.
    }
    \label{fig:DGCluster}
\end{figure}

The case we discuss here involves six DGCs, all located within approximately 1\degr ~of the field of the galaxies PGC 043894, PGC 1070195, and PGC 183425 (see Figure \ref{fig:DGCluster}). Table \ref{tab:darkCluster} reports the names of these sources, the $cz_{\rm \odot, radio}$, the integrated flux (S$_{\rm sum}$), the W$_{50}$, and the hydrogen mass. It is worth noting that PGC~043894, PGC~1070195, and PGC~183425 lie at 98 Mpc ($z=0.0228$), 125 Mpc ($z=0.0292$), and at 200 Mpc ($z = 0.0466$), respectively, and are thus background galaxies, not associated with the DGCs (all located within 24 Mpc); PGC 043894 was correctly identified by FAST, which lists a position and a velocity compatible with the optical ones.

\begin{table*}
 	\centering
 	\caption{The six DGCs around the galaxies PGC 043894, PGC 1070195, and PGC 183425. The table columns list the $cz_{\odot, \mathrm{radio}}$, the integrated flux density over the line profile, W$_{50}$, and the logarithm of the \ion{H}{i} mass. The sources are sorted in right ascension so that they can be easily spotted in Figure \ref{fig:DGCluster} from right to left.}
 	\label{tab:darkCluster}
 	\begin{tabular}{ccccc}
 		\hline
 		\textbf{Name} & $\mathbf{cz_{\odot,\mathrm{radio}}}$ \textbf{[km s$^{-1}$]} & \textbf{S$_{\mathrm{sum}}$ [mJy km s$^{-1}$]} & \textbf{W$_{50}$} \textbf{[km s$^{-1}$]} & \textbf{$\mathbf{\log (M_{HI} / M_{\odot})}$}\\
 		\hline
 		J125312.52-034857.7 & 1529.58 $\pm$ 1.23 & 622.27 $\pm$ 93.7 & 29.54 $\pm$ 2.46 & 8 $\pm$ 0.07\\
 		J125358.60-034901.6 & 1531.93 $\pm$ 2.37 & 345.36 $\pm$ 55 & 48.06 $\pm$ 4.75 & 7.67 $\pm$ 0.08\\
 		J125417.40-035248.2 & 1500.41 $\pm$ 1.88 & 310.7 $\pm$ 47.4 & 49.32 $\pm$ 3.75 & 7.65 $\pm$ 0.07\\
            J125427.18-034603.6 & 1503.11 $\pm$ 1.39 & 622.67 $\pm$ 84.72 & 25.90 $\pm$ 2.78 & 7.94 $\pm$ 0.07\\
            J125542.52-032925.3 & 1488.28 $\pm$ 0.51 & 697.86 $\pm$ 37.72 & 25.97 $\pm$ 1.02 & 7.9 $\pm$ 0.05\\
            J125557.87-033529.1 & 1485.48 $\pm$ 2.02 & 255.53 $\pm$ 46.64 & 29.75 $\pm$ 4.03 & 7.56 $\pm$ 0.08\\
 		\hline
 	\end{tabular}
 \end{table*}

The integrated flux density over the line profile (S$_{\mathrm{sum}}$) changes between sources, but the systemic velocity $cz_{\odot,\mathrm{radio}}$ is remarkably similar, within $\sim 50$ km s$^{-1}$. The inferred distances are similar too, with the closest source (J125557.87-033529.1) located at 22.99 $\pm$ 1.15 Mpc, while the most distant (J125358.60-034901.6) is located at 23.72 $\pm$ 1.19 Mpc. If we consider a mean distance of $\sim 23.4$ Mpc, the sources are distributed over a total extent of approximately 300 kpc, with individual separations between sources ranging from about 80 to 100 kpc. The inferred \ion{H}{i} mass for each source is between $\log (M_{\mathrm{HI}}/M_{\odot})\sim 7.6$ and $\log (M_{\mathrm{HI}}/M_{\odot})\sim 8$, leading to a total mass (adding up all \ion{H}{i} of the six sources) of about $\log (M_{\mathrm{HI}}/M_{\odot})\sim 8.6$.

From a purely statistical point of view, the detection of as many as six DGCs within less than one square degree is highly unlikely. The average density of our 70 DGCs is 0.0097 per deg$^{2}$ (see Section \ref{subsec:skyDistribution}), so within the area shown in Figure \ref{fig:DGCluster}, which covers 0.6 deg$^{2}$, we would expect to find only 0.006 DGCs on average. Assuming a Poisson number statistics, the probability of detecting six DGCs in this area is about 10$^{-17}$, making it virtually impossible that all these sources are real. A possible explanation is an RFI when this part of the sky was observed, and/or the DGCs are not Poisson distributed over the sky. We conservatively excluded these six sources from our DGC catalogue.

\subsection{The detection near UGC 4107}\label{subsec:ugc4107}
\begin{figure}
    \includegraphics[width=\columnwidth]{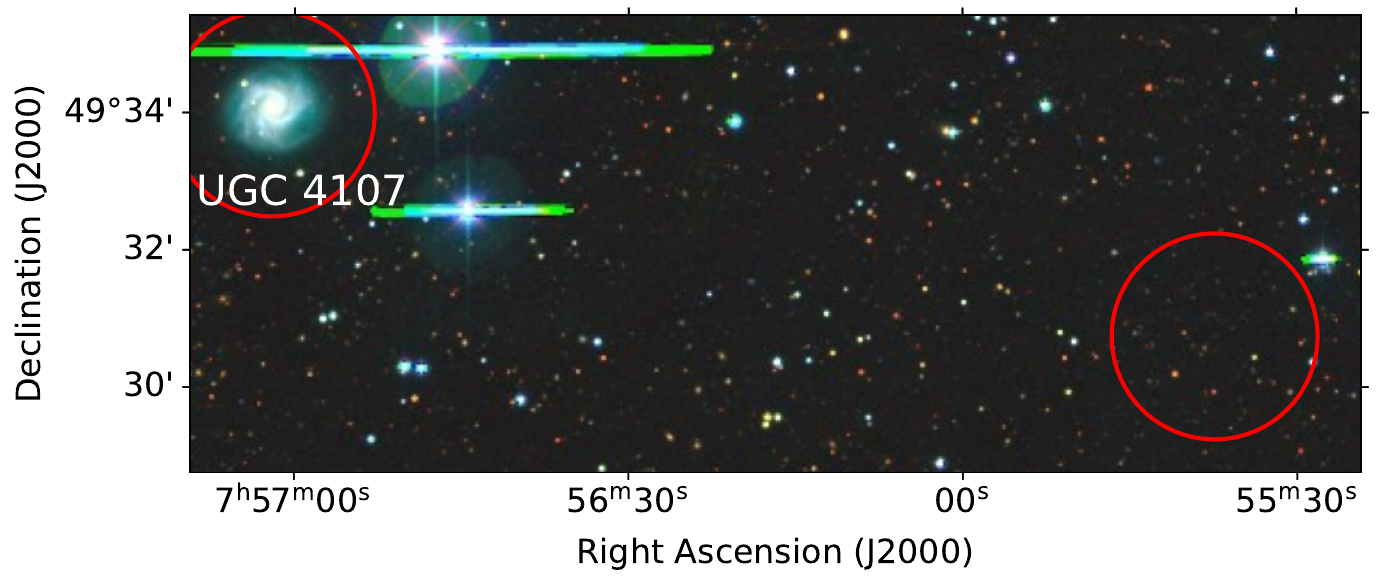}
    \caption{The FASHI detection near UGC 4107. The two red 1\farcm5 ~radius circles highlight the FASHI detections in this region. UGC 4107 was not detected by ALFALFA, and the $cz_{\odot, \mathrm{radio}}$ and the W$_{50}$ from FAST are compatible with those reported by \citet{2018A&A...609A..17J}. The other FAST detection is approximately 21\arcmin \ south-west of UGC 4107. If projected at the distance of $\sim 48$ Mpc, this angular separation corresponds to a linear distance of about 300 kpc.}
    \label{fig:ugc4107}
\end{figure}

The DGC J075537.39+493044.5 is located near the galaxy UGC 4107, which was correctly detected by FASHI (see Figure \ref{fig:ugc4107}, which shows both the \ion{H}{i} detection positions for UGC 4107 and the DGC). The FASHI detection of UGC 4107 is plausible, as the measured properties are consistent with those previously reported by \citet{2018A&A...609A..17J} (see Table \ref{tab:ugc4107}). It is assigned a distance of 48 Mpc in FASHI.

\begin{table}
    \centering
    \caption{Radial velocity, linewidth, and peak flux of UGC 4107 as measured by FASHI and by \citet{2018A&A...609A..17J}}
    \label{tab:ugc4107}
    \begin{tabular}{lccc}
        \hline
        & $cz_{\odot}$ [km\,s$^{-1}$] & $W_{50}$ [km\,s$^{-1}$] & ${\rm F_P}$ [mJy] \\
        \hline
        \textbf{FASHI}       & $3452.16 \pm 1.64$ & $141.84 \pm 1.19$ & $31.80 \pm 2.84$ \\
        \textbf{Jones et al.} & $3512.4 \pm 7.7$   & $153.5 \pm 10.9$  & $32.1 \pm 3.1$ \\
        \hline
    \end{tabular}
\end{table}

However, the dark \ion{H}{i} source was not detected by previous \ion{H}{i} surveys and shows atypical values in terms of line shape. In particular, W$_{50} = 5.65 \pm 3.28$ km s$^{-1}$ is suspiciously small. Given the channel spacing of 1.6 km s$^{-1}$ at 1420 MHz in the FASHI catalogue, and 6.4 km s$^{-1}$ after smoothing, this source is spectrally unresolved. On the other hand, W$_{20}$ is quite large, with a value of $49.25 \pm 4.92$ km s$^{-1}$. 

The value $cz_{\odot,\mathrm{radio}} = 3452.16 \pm 1.64$ km s$^{-1}$ is remarkably similar to the recessional velocity of UGC 4107, and the source lies over 300 kpc away from UGC 4107 in projection, assuming a distance of about 48 Mpc. If this source were associated with UGC 4107, it would be one of the most remote tidal tails or debris features ever identified around a galaxy \citep[][]{2012MNRAS.419L..19S, 2020arXiv200207312K}. Given these considerations, we conservatively decided to exclude this source from the final validated DGC catalogue.

\subsection{The detection near PGC 2160003}\label{subsec:pgc2160003}
\begin{figure}
    \includegraphics[width=\columnwidth]{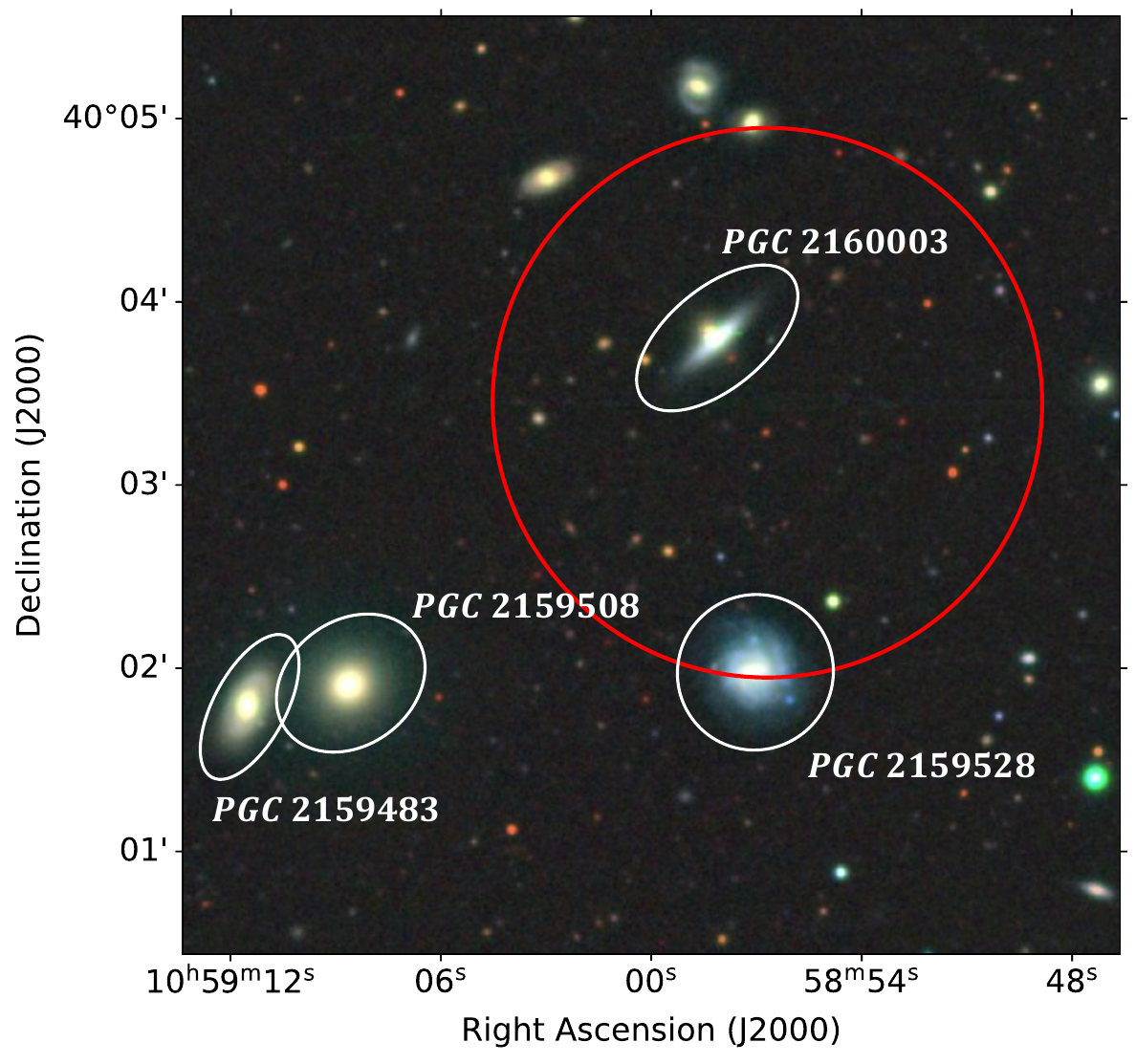}
    \caption{Legacy image showing the galaxies PGC 2160003, PGC 2159528, PGC 2159508, PGC 2159483 (highlighted by white ellipses). All these galaxies are listed in the Siena Galaxy Atlas, but were filtered out since their recessional velocities are above the threshold of $cz_{\odot} = 3500$ km s$^{-1}$. The FASHI detection is in the centre of the 1\farcm5 radius red circle. Note that PGC 2160003 is probably two different galaxies aligned along the line of sight.}
    \label{fig:pgc2160003}
\end{figure}

The source J105856.69+400327.1 (see Figure \ref{fig:pgc2160003}) is located near the galaxies PGC 2160003 (optical $cz_{\odot}\simeq 45270$ km s$^{-1}$), PGC 2159528 (optical $cz_{\odot}\simeq 7500$ km s$^{-1}$), PGC 2159508 (optical $cz_{\odot}\simeq 17320$ km s$^{-1}$), and PGC 2159483 (optical $cz_{\odot}\simeq 17200$ km s$^{-1}$), all listed in the SGA. These galaxies are therefore in the background, and were filtered out due to their too high radial velocities in the first step of Figure \ref{fig:flowchart}. The FASHI source has $cz_{\odot, \mathrm{radio}}=3082.6 \pm 1.2$ km s$^{-1}$; therefore, it was considered a perfectly legitimate DGC. However, an examination of Figure \ref{fig:pgc2160003} reveals an interesting detail that is worth a brief discussion. PGC 2160003 is located almost at the centre of the beam; however, it does not appear to be a single galaxy, but rather two galaxies at different redshifts, superimposed along the line of sight due to projection effects. In particular, it is possible to spot an orange halo, beyond a more pronounced blue halo in front of it. It is therefore possible that the radial velocity was calculated from a spectrum resulting from a blend of the light of the two galaxies, thus leading to an incorrect value. The galaxy spectrum is available in the HyperLEDA database, confirming the superposition of two different galaxies. In particular the overall spectrum is that of an elliptical galaxy, with absorption lines redshifted accordingly to $cz_{\odot}\sim$ 45270 km s$^{-1}$, but with some emission lines (such as H$\alpha$ and [O III]) at $cz_{\odot}\sim3000$ km s$^{-1}$, which in turn is compatible with the \ion{H}{i}-measured $cz_{\odot, \mathrm{radio}}$. We conclude that the spectrum is a blend of two different sources at different radial velocities, and the DGC might be physically associated with the foreground blue galaxy, which has been mistakenly assigned a $cz_{\odot} = 45268 \pm 7$ km s$^{-1}$. Given this explanation, we exclude it from our final DGC catalogue.

\subsection{The three sources with very low W$_{50}$}\label{subsec:lowW50}
Other sources that raise some concerns are three detections showing extremely low W$_{50}$ values, below 10 km s$^{-1}$. One of these sources has already been discussed in Section \ref{subsec:ugc4107}, the detection near UGC 4107. The other two (J102341.54+502450.0 and J001847.80-003030.6) share similar characteristics, namely a very low W$_{50}$, and W$_{20}$ between 50 and 60 km s$^{-1}$.
Although dark galaxies are indeed expected to have very low W$_{50}$ values, the spectral resolution of the FASHI catalogue (6.4 km s$^{-1}$) does not allow the \ion{H}{i} emission peak to be spectroscopically resolved in these cases, making it difficult to place full confidence in such detections. Moreover, galaxies generally show consistent W$_{50}$ and W$_{20}$ values, as the wings of the \ion{H}{i} profiles are typically quite steep. The large discrepancy between W$_{50}$ and W$_{20}$ in these sources instead suggests that they may be residual RFI, or possibly OH megamasers \citep[powerful line emissions at redshifts 0.176–0.186 from merging galaxies,][]{2002AJ....124..100D}. With the exception of the source near UGC 4107, the other two are isolated, thus ruling out a tidal origin. All three were excluded from the final catalogue.

\subsection{Description of the final DGC catalogue}\label{subsec:DGCcatalogue}
Table \ref{tab:DGC_example} reports an example of the validated Dark Galaxy Candidates catalogue. The full list is available in the online Supplementary materials with additional columns. Here we list the principal columns such as the name of the source, the \ion{H}{i} position in the sky, the recessional velocity computed from \ion{H}{i}, the W$_{50}$ and W$_{20}$, the flux peak, the S/N, the distance, and the logarithm of the \ion{H}{i} mass. These columns are listed directly from values in the FASHI catalogue \citep[see][for a detailed description of the various columns]{2024SCPMA..6719511Z}. We also added here a column listing the expected radius of the \ion{H}{i} disc inferred by the \ion{H}{i} mass using the relation described by \citet{2016MNRAS.460.2143W}. In the online catalogue, we added a column with comments we think are useful to assist future studies and facilitate a more informed selection of sources. In this latter column, we flagged the three sources (4\% of the catalogue) that present a foreground star within 30\arcsec \ of the beam centre that is so bright as to make the visual assessment more difficult.

\begin{table*}
\centering
\caption{The Dark Galaxy Candidates catalogue. The full table is available in the online material with additional columns.}
\label{tab:DGC_example}

\begin{tabular}{ccccccccccc}
\toprule
Name & RA & Dec & $cz_{\odot}$ & W$_{50}$ & W$_{20}$ & 
F$_{\rm peak}$ & S/N & distance & 
$\log (M_{HI}/M_{\odot})$ & R$_{\rm HI}$ \\
 & $[\degr]$ & $[\degr]$ & [km s$^{-1}$] & [km s$^{-1}$] & [km s$^{-1}$] & [mJy] & & [Mpc] & &[kpc]\\
\midrule
J032929.05-032610.7 & 52.371 & -3.4363 & 2006.75 & 
114.71 & 118.19 & 2.68 & 8.37 & 27.44 & 7.6 & 1.78 \\

J124838.01+500346.8 & 192.1584 & 50.063 & 2230.9 & 15.78 & 61.07 & 13.08 & 16.87 & 35.69 & 8.11 & 3.23 \\

J084333.16+452205.5 & 130.8882 & 45.3682 & 1121.16 & 90.24 & 117.33 & 8.09 & 13.36 & 21.45 & 7.9 & 2.53 \\

J043131.94-030035.1 & 67.8831 & -3.0097 & 1921.66 & 135.83 & 153.96 & 4.09 & 12.03 & 26.43 & 7.95 & 2.68 \\

J110735.21+313821.8 & 166.8967 & 31.6394 & 3045.12 & 93.74 & 108.79 & 5.36 & 11.57 & 48.48 & 8.25 & 3.81 \\

\vdots & \vdots &\vdots &\vdots &\vdots &\vdots &\vdots &\vdots &\vdots &\vdots &\vdots \\
\bottomrule
\end{tabular}

\end{table*}

\section{Discussion}\label{sec:discussion}
In this section, we will discuss the properties of the 70 validated DGCs, comparing them with the properties of the sources with optical counterparts (listed both in the optical catalogues and the new sources from the `O' class, i.e. a total of 2690 sources).

\subsection{Distribution over the sky and the abundance of DGCs}\label{subsec:skyDistribution}

Figure~\ref{fig:DGCMoll} shows the positions of the 70 DGCs and 2690 \ion{H}{i} sources with an optical counterpart across the sky using a Mollweide projection.
The sources are confined to two stripes, one near the equator, and the other one between 30 and 60 degrees in declination, due to the limited sky coverage of the FASHI first-release catalogue.

Examining the spatial distribution of the DGCs and optical galaxies from FAST across the sky, they appear to have a similar distribution, although our sample of DGCs is relatively small in number and limited in distribution on the sky. In order to address this, we performed a 2D Kolmogorov-Smirnov (KS) test as described by \citet{1987MNRAS.225..155F}, obtaining a {\it p-}value of 0.03. Even though this result could be regarded as an indication of a slightly different spatial distribution between the DGCs and the galaxies with an optical counterpart, the statistically small number of DGCs does not allow a definitive conclusion. The similar distribution over the sky may be seen as a tension with what has been predicted by \citet{2024ApJ...962..129L}, who claimed that dark galaxies tend to reside in low-density environments. However, given the limitations mentioned above, a more quantitative analysis will have to wait until the full FASHI catalogue is available.

Based on the detections we have, it is possible to estimate the surface density of DGCs per square degree and infer the total number of dark galaxies within $\sim$50 Mpc that could be observed across the entire sky, or within the portion of the sky observable by the FAST radio telescope. First, we evaluated the total area covered by the first release of the catalogue, which amounts to approximately 7245 deg$^2$, corresponding to about 17.6\% of the whole sky. Considering 70 DGCs, this yields a surface density of 0.0097 DGC/deg$^2$, or, in other words, roughly one DGC every $\sim 103$ square degrees. Extrapolating this density to the entire sky, we expect a total of about 400 DGCs within 50 Mpc. If we restrict the estimate to the sky observable by FAST, the expected number of DGCs is approximately 200. 

However, we emphasise that these numbers are only approximate. Our selection process was rather strict, and some potential DGCs may have been excluded. 

As a final note, it is interesting to estimate the `yield' of our filtering procedure, evaluating the detection rate of DGCs to the total of sources with $cz_{\odot, \rm radio}\leq 3500$ km s$^{-1}$. The detection of 70 DGCs out of 3436 velocity-filtered sources from the FASHI catalogue corresponds to a percentage of 2\%. In the case of uncatalogued optical galaxies, the rate is approximately 15.5\%. 

\subsubsection{Galactic extinction}

\begin{figure*}
    \includegraphics[width=\textwidth]{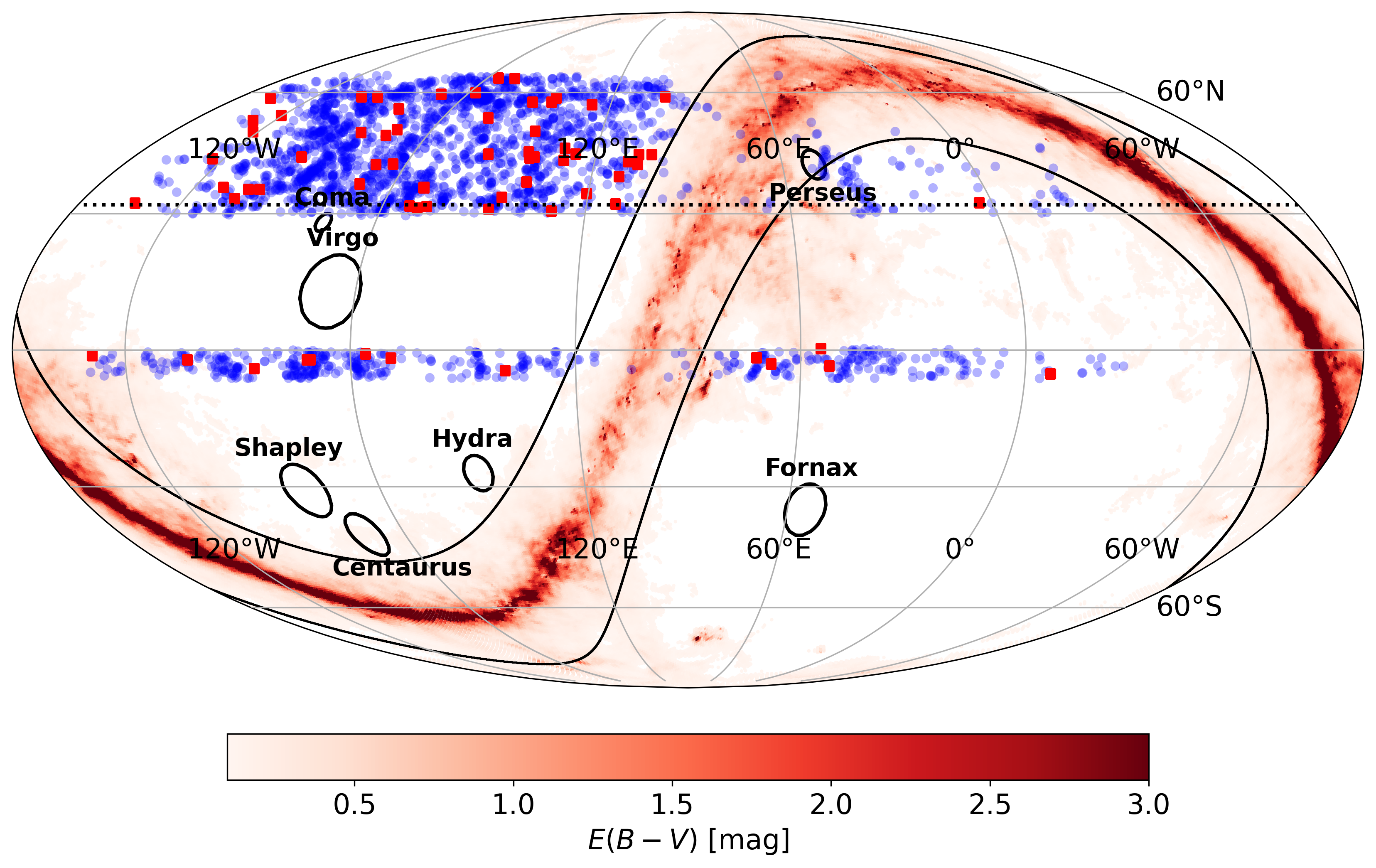}
    \caption{Distribution over the sky of the DGCs (D; red squares), and the galaxies with an optical counterpart (both found in optical catalogues and the `O' class visually inspected). The red colourmap shows the E(B -- V) Schlegel, Finkbeiner, and Davis dust map \citep[SFD, ][]{1998ApJ...500..525S} corrected for extragalactic emission by \citet{2023ApJ...958..118C}. 
    The black lines encompass the Galactic plane between $b = -15$ and $b = +15$. To put the DGC positions in context, the black ellipses highlight well-known galaxy clusters. The dotted line at Dec = 32\degr indicates the separation between MzLS+BASS (northern part) and DECaLS (southern part). At a first qualitative visual inspection, the DGCs do not show any particular difference in the distribution over the sky with respect to the optical galaxies.
    }
    \label{fig:DGCMoll}
\end{figure*}

The DESI Legacy DR10 does not cover the regions near the Galactic plane between $|b|\sim 18\degr$, so all the \ion{H}{i} detections that fall within these limits were classified as `Unsure' (class `X') since we do not have optical images to perform the visual assessment. Therefore, most DGCs are `intrinsically' selected to be located in low-extinction regions (see Figure \ref{fig:DGCMoll}).
Clearly, it is possible that some \ion{H}{i} detections located near the Galactic plane could be DGCs, but the overcrowding by the stars, as well as the high Galactic extinction, would make the optical assessment extremely challenging and prone to misidentifications.
We used the dust map by \citet{2023ApJ...958..118C} to evaluate the E(B--V) extinction for our selection of DGCs, and the mean value is 0.03 mag. Given the low mean extinction to DGCs, Galactic dust does not explain the lack of an optical counterpart.

From this perspective, these sources are well-suited for deep optical integrations, which can further constrain the upper limits on their total stellar mass.

\subsection{Observed properties of the DGCs}\label{subsec:DGCprop}
\begin{figure*}
    \includegraphics[width=0.8\textwidth]{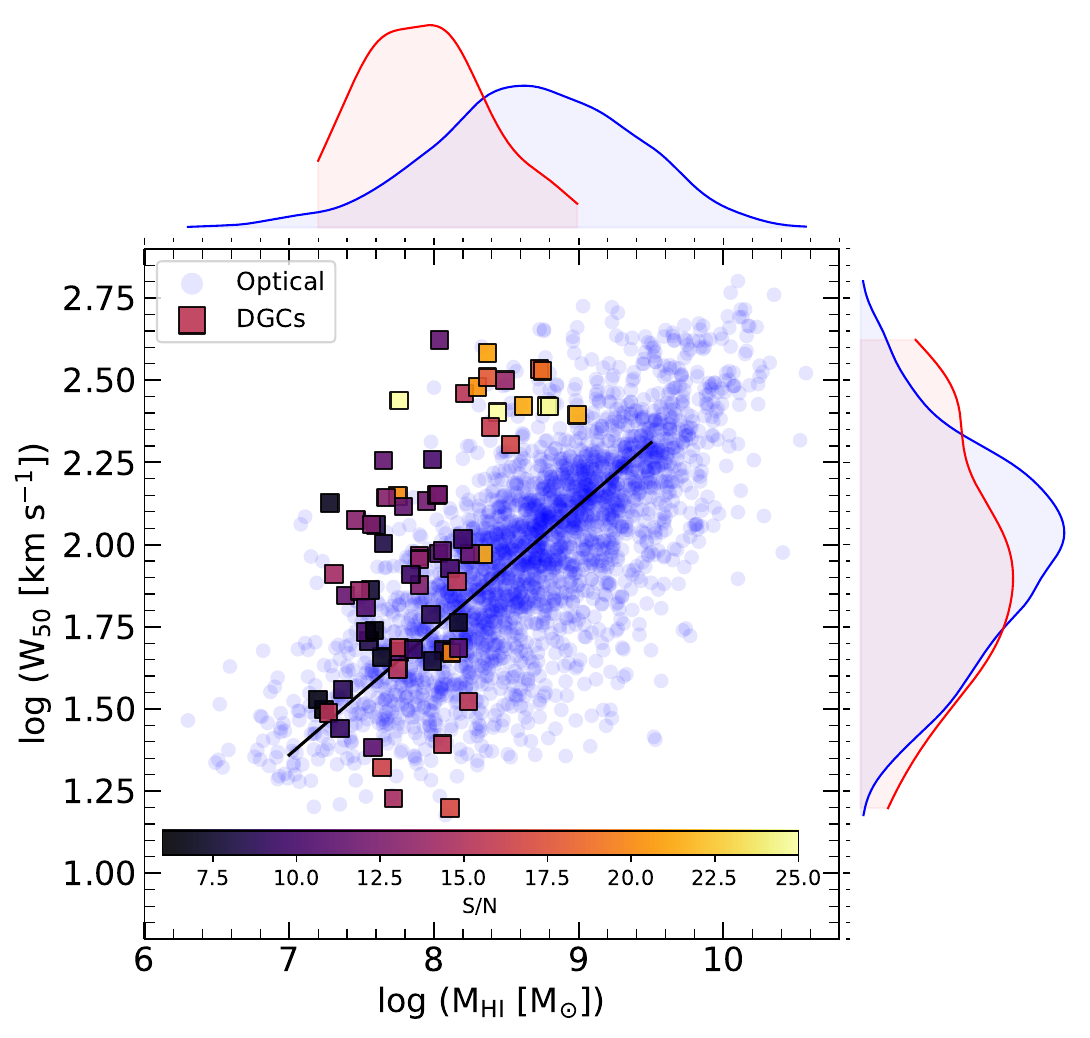}
    \caption{Plot of the \ion{H}{i} velocity width (W$_{50}$) versus the inferred hydrogen mass M$_{\mathrm{HI}}$ for \ion{H}{i} sources with an optical counterpart (blue circles) and for our DGCs (squares, colour-coded by S/N). On the top and right, the probability density functions of the individual quantities are also plotted (red for the DGCs, blue for the sources with an optical counterpart). The solid black line highlights the scaling relation of W$_{50} \propto M_{\mathrm{HI}}^\alpha$, where $\alpha = 0.38$, as reported by \citet{2022MNRAS.511.2585D}. 
    The scatter is partly due to the fact that, lacking information on inclination, we cannot recover $V_{\mathrm{rot}}$. Despite the scatter, DGCs follow a different distribution from optical galaxies.} 
    \label{fig:W50_MHI_DGCs_opt}
\end{figure*}

To better put the DGCs in context, we analysed their distribution in the W$_{50}$ -- M$_{\rm HI}$ plane, and compared that with the distribution of the \ion{H}{i} sources with an optical counterpart (both previously catalogued and from the `O' class), as shown in Figure \ref{fig:W50_MHI_DGCs_opt}.

The galaxies with an optical counterpart exhibit a trend which is very similar to what was found by \citet{2022MNRAS.511.2585D}, namely W$_{50} \propto M_{\rm HI}^\alpha$, with $\alpha = 0.38$ (black line). In \citet{2022MNRAS.511.2585D} study, this trend is followed by low-mass, blue, star-forming galaxies, and thus is similar to our sample of optical galaxies. This trend is not followed by our DGCs.

The histograms located at the top and right of the plot show the distribution of the \ion{H}{i} mass and the W$_{50}$, respectively, using the same colour code as the main plot. 
The distributions are normalised to illustrate the differences better, as the datasets contain different numbers of sources. 
 
In this context, the W$_{50}$ distribution of our DGCs shows both higher and lower values, with no statistically significant difference relative to the distribution of galaxies with optical counterparts, as confirmed by a KS test. On the other hand, the distribution of \ion{H}{i} masses is different between the two classes, with the DGCs showing a statistically significantly lower \ion{H}{i} mass compared with the optical counterparts.

The overall distributions of our DGC sample and optical galaxies in the W$_{50}$ -- M$_{\rm HI}$ plane are slightly different: the DGCs appear to follow a much steeper trend compared to some of the optical galaxies. However, at high W$_{50}$, they seem to follow a different trend. We speculate that some of the very high W$_{50}$ measurements may arise from multiple HI clouds along the line-of-sight blending together.

Recently, \citet{2025ApJS..281...66C} analysed low \ion{H}{i} mass (< $10^8$ M$_{\odot}$) optical dwarf galaxies in the FASHI catalogue, in order to better understand the dynamical state of these systems and link the \ion{H}{i} content with the stellar mass, inferred by images from the Legacy.
Their selected galaxies are morphologically very similar to our optical counterparts, i.e. blue irregular dwarf galaxies, validating our selection procedure (see their figures C1 and C2). \citet{2025ApJS..281...66C} discarded several \ion{H}{i} sources without a clear optical counterpart as they are not within the scope of their study. We speculate that these discarded sources (those within 50 Mpc and M$_{\rm HI} < 10^8$ M$_{\odot}$) may be consistent with the detections we have classified as DGCs.
However, since we do not have the list of the discarded sources, we cannot compare in detail our sample with their work.

In Figure \ref{fig:W50_MHI_DGCs_opt} we colour-coded the DGCs by the S/N of the FAST detection. We remember here that \citet{2024SCPMA..6719511Z} applied, for the compilation of the FASHI catalogue, a S/N threshold of 5, to ensure the most reliable sources. In our case, the lowest S/N we have for a DGC is $\sim 6$. To put this value in context, despite a low S/N of $\sim 7$, an \ion{H}{i} detection was successfully cross-matched with a galaxy in an optical catalogue (sharing the same radial velocity). Only three of our DGCs have an S/N lower than 7. Apart from the DGCs with high W$_{50}$, which show slightly higher S/N than the rest of the sources, there is no clear trend of S/N within the distribution.

\begin{figure}
    \includegraphics[width=\columnwidth]{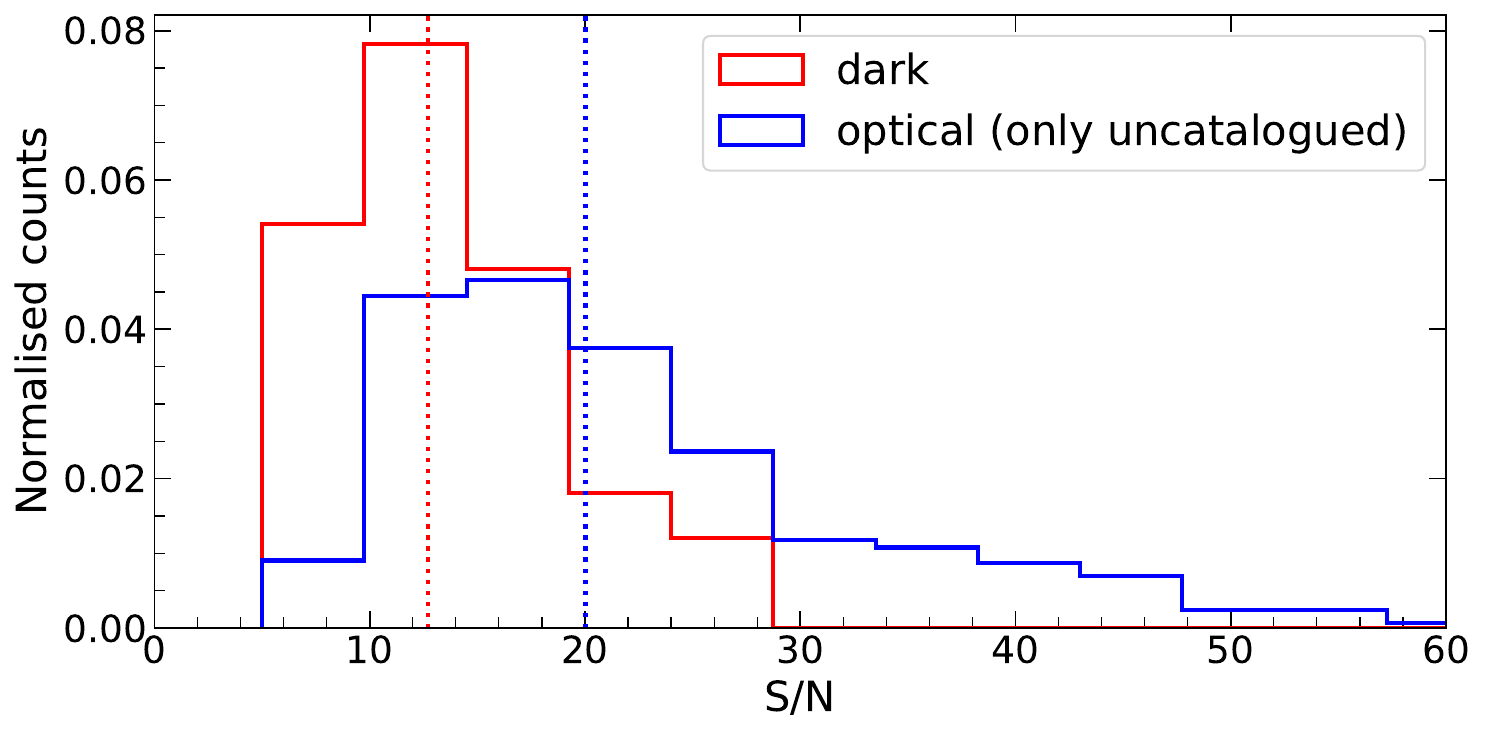}
    \caption{The distributions of the S/N for the \ion{H}{i} detections with an optical counterpart (blue line, here we report only the previous uncatalogued optical galaxies) and the DGCs (red line). The vertical dotted lines indicate the median for the respective distributions (20.0 for the optical counterparts and 12.7 for the DGCs). The counts have been normalised for a better visualisation. As expected by their systematically lower \ion{H}{i} masses, the DGCs show a systematically lower S/N.} 
    \label{fig:SNRdistr}
\end{figure}

Figure \ref{fig:SNRdistr} shows the distribution of the S/N for the DGCs (red line) and the \ion{H}{i} detections with an optical counterpart (blue line), with their median value highlighted by a dotted vertical line of the same colour. In this case, we considered only the potentially uncatalogued optical sources (the `O' class), i.e. excluding the \ion{H}{i} sources that were successfully cross-matched with catalogues of known galaxies. In fact, the successfully cross-matched sources show, on average, a very high S/N ratio given their \ion{H}{i} mass, thereby saturating the high-S/N end of the distribution. On the other hand, the uncatalogued sources show \ion{H}{i} detections that are more compatible with our DGC ensemble.

The DGCs show a systematically lower S/N compared with optical counterparts, as indicated by the median of the distributions; for the DGCs, the median S/N is 12.7, whereas for the optical counterparts, it is 20.0. This is expected, since the \ion{H}{i} mass for the DGCs is systematically lower than the optical counterparts.

\subsection{Comparison with other observational samples}\label{subsec:compK25}

In this section, we compare the distribution in the W$_{50}$--M$_{\rm HI}$ plane of our DGCs with the dark galaxy candidates from the ALFALFA survey described by \citealt{2025ApJS..279...38K} (hereafter K25) and from the WALLABY pilot survey by \citealt{2025PASA...42...87O} (hereafter OB25). K25 selected, from a preliminary catalogue of 344 \ion{H}{i} sources without optical counterparts, 142 dark galaxy candidates. The procedure followed by K25 is similar to that used in this work, i.e., utilising the DESI Legacy and the NED database to further refine the catalogue, excluding potential optical counterparts and tidal tails. The criteria used are slightly different from what we used: the velocity difference was set to be smaller than 400 km s$^{-1}$ (we used a velocity tolerance of $\pm$ 200 km s$^{-1}$); to avoid tidal tails, they set a fiducial distance of 75 kpc (we used 150 kpc). We also note that the dark galaxy candidates selected by K25 are not limited in distance.

Since the visual detection was performed manually, as we did, the outcome for the same source can differ between our study and the work of K25. In Appendix \ref{app:optCountQuest} we further discuss this point.

The approach followed by OB25 is similar too. They examined the optical counterparts of more than 1800 \ion{H}{i} sources in the pilot fields of WALLABY using the DESI Legacy and found that 25 sources lack optical counterparts. We remind the reader here that WALLABY is an interferometric survey made by the Australian SKA Pathfinder telescope, therefore the beamsize is much smaller ($\sim30\arcsec$), but with a lower sensitivity ($\sim 1$ dex in \ion{H}{i} mass less than FAST).

\begin{figure*}
    \includegraphics[width=0.8\textwidth]{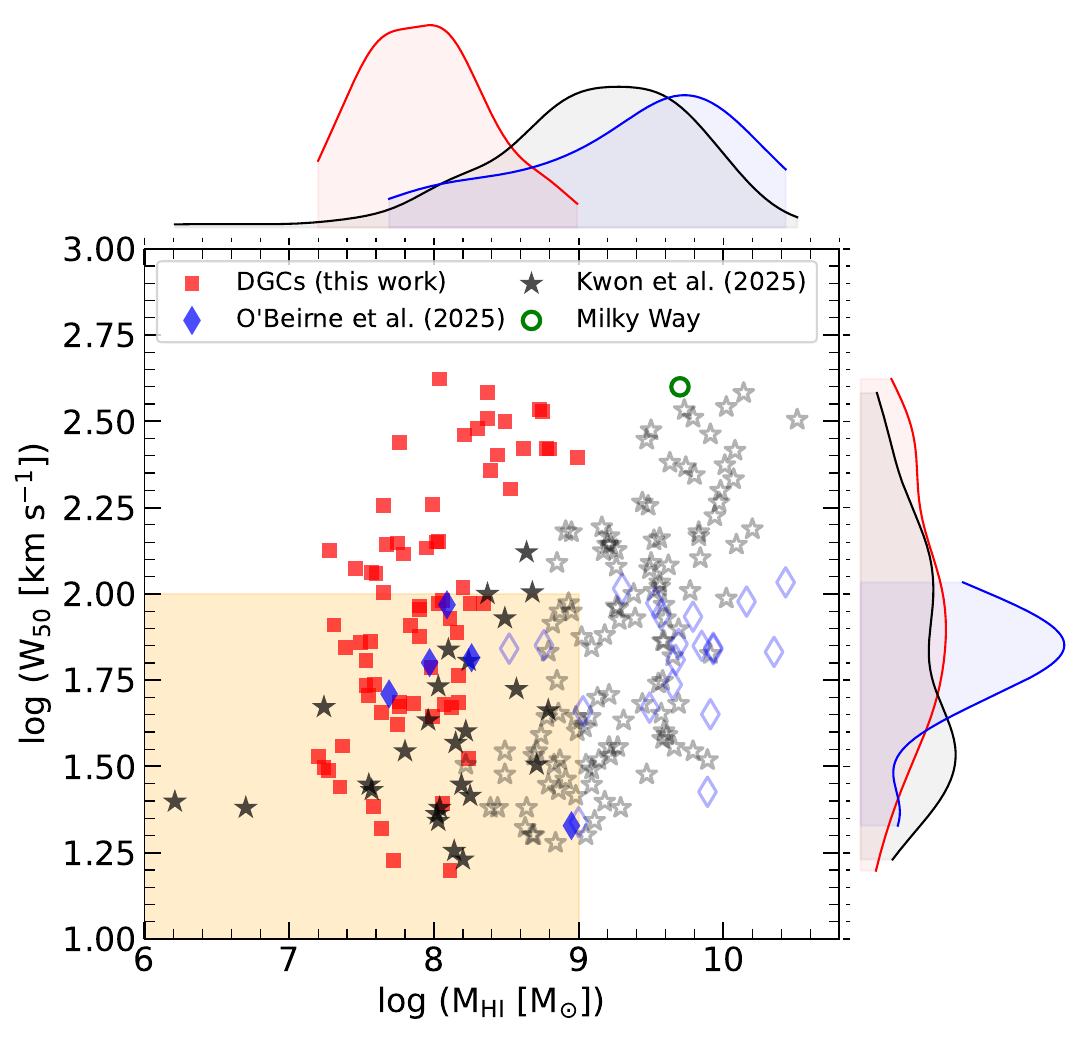}
    \caption{The distribution of our DGCs, those of \citet{2025ApJS..279...38K}, and those of \citet{2025PASA...42...87O} in the W$_{50}$--M$_{\rm HI}$ plane. For reference, we also indicate the position of the Milky Way, adopting the W$_{50}$ it would have if viewed edge-on (green circle). The histograms on the top and on the right represent the probability distribution functions of the single quantities. The candidates from K25 and OB25 within 50 Mpc are highlighted by solid symbols. The orange-shaded rectangle highlights the region of the plane compatible with the theoretical predictions for dark galaxies as discussed by \citet{2006MNRAS.368.1479D}. Several DGCs from our work, K25, and OB25 are in tension with the Davies theoretical model.} 
    \label{fig:DCGKwon}
\end{figure*}

Figure \ref{fig:DCGKwon} shows the distribution of our DGC sample (red squares), the K25 candidates (black stars), and the OB25 candidates (blue diamonds) in the W$_{50}$--M$_{\rm HI}$ plane. We note that a major difference between the three samples is that, in our case, we limited ourselves to sources within 50 Mpc, whereas K25 and OB25 imposed no distance constraint. The K25 and OB25 candidates located within 50 Mpc are plotted as solid symbols, whereas the candidates beyond 50 Mpc are plotted as empty symbols. The orange-shaded rectangle represents the region of the plane compatible with the theoretical predictions by \citet{2006MNRAS.368.1479D}, which we discuss in detail in the next section.

The distribution of our sample differs from that of the OB25 sample, as our candidates show a systematically lower mass. This could be explained by the different sensitivities of the two surveys. If we look at W$_{50}$, our sample covers a wide range of values, whereas the OB25 sample appears to be limited to $\log W_{50}\lesssim 2$.

Also, in the case of the K25 sample, the distribution looks quite different from our sample in the W$_{50}$--M$_{\rm HI}$ plane; even if they share similar distributions in W$_{50}$, the \ion{H}{i} mass distributions are substantially different, confirmed by a KS test. It is important to note that the \ion{H}{i} mass is not a directly measured quantity, but rather it is derived based on various assumptions about distance and the integrated flux ($\propto D^2 S$). However, if we consider only the sources within 50 Mpc in the K25 sample, the two distributions are less in tension, and even our high-W$_{50}$ sources have no analogues in their DGCs distribution. The two K25 sources with the lowest \ion{H}{i} mass, below 10$^7$ M$_{\odot}$, exhibit radial velocities of $\lesssim$ 500 km~s$^{-1}$, so it is possible that they could be local High Velocity Clouds \citep[HVC,][]{1999ApJ...514..818B, 2001ApJS..136..463W}.   

The number of \ion{H}{i} sources detected below 10$^{9}$ M$_{\odot}$ is greater in the case of the FASHI catalogue compared with the ALFALFA survey. This can be explained by the fact that FAST is more sensitive than the Arecibo telescope by a factor of $\sim 3$, so we are expecting to detect \ion{H}{i} sources with lower masses.

An interesting exercise is to estimate the fraction of dark galaxies out of all sources from the two samples. From FASHI, we started with 3436 sources, and we identified 70 DGCs within 50 Mpc, giving a fraction of $\sim$2\%. In order to make a fair comparison with ALFALFA, we need to consider only the sources within 50 Mpc. Out of 3560 ALFALFA sources within 50 Mpc, K25 identified 26 DGCs, or $\sim$1\%.

Several sources from both the K25 and OB25 samples show an \ion{H}{i} mass higher than what is expected for dark galaxies (see the next Section). Considering 10$^{-2}$--10$^{-3}$ as a rough estimation for the M$_{\rm HI}$/M$_{\rm halo}$ ratio, we would expect a halo mass of 10$^{11}$--10$^{12}$ M$_{\odot}$ for these systems, so they should form stars, albeit they do not have a clear optical counterpart. In Figure \ref{fig:DCGKwon} we reported, as a green circle, the position of the Milky Way in the W$_{50}$ -- M$_{\rm HI}$ plane. Several galaxies from the K25 and OB25 samples show an \ion{H}{i} mass larger than that of the Milky Way.

Using the tight relation $\log R_{\rm HI} \propto 0.506 \log M_{\rm HI}$ described by \citet{2016MNRAS.460.2143W}, which appears to be independent of the morphology of the galaxy, we can infer that the K25 and OB25 dark galaxy candidates are systematically bigger in size with respect to our sample.

The overall tension between our sample and K25 can also be alleviated if some of the K25 sources are tidal tails or plumes stripped from nearby galaxies. \citet{2018MNRAS.477.2741Y} analysed the case of AGC 249460, which is in the vicinity of NGC 5577, and which is included as a dark galaxy candidate in K25. Given the small velocity difference with NGC 5577, the morphology and the kinematics of the system, AGC 249460 is consistent with \ion{H}{i} material stripped from NGC 5577. In our selection process, this source could have been classified as a tidal tail and excluded from the final catalogue of DGCs.

\subsection{Comparison with theoretical predictions}\label{subsec:theopred}
We now discuss the distribution of DGCs in the W$_{50}$--M$_{\rm HI}$ plane in light of the results from simulations.
\citet{2006MNRAS.368.1479D} studied the existence and the properties of simulated dark galaxies in the context of the first \ion{H}{i} extragalactic survey, and to put into context the recent discovery of VIRGOHI21 \citep[][]{2004MNRAS.349..922D, 2005ApJ...622L..21M}. They used an analytical model to produce stable gaseous rotating discs embedded within DM haloes, of which they have different properties, such as the total mass, the gas column density and the circular velocity. Their dark galaxies have lower gas surface densities and higher spins, which prevent star formation. In Figure \ref{fig:DCGKwon}, the orange-shaded rectangle encompasses the region of the plane that should be populated by the dark galaxies simulated by \citet{2006MNRAS.368.1479D}, remembering that this sector should be regarded as a `maximum probability' region. For the \ion{H}{i} masses, \citet{2006MNRAS.368.1479D} predict $\sim 20\%$ of the dark galaxies have masses above 10$^{7}$ M$_{\odot}$, while the others lie below this limit. In light of this, and given the sensitivity of the FAST telescope, we are probing the high-mass end of the distribution. This result was recently confirmed by \citet{2026arXiv260104024G}, who used the HESTIA and NIVARIA-LG simulations to study a population of dark galaxies in the Local Group, finding that dark galaxies without stars have an \ion{H}{i} mass below 10$^{8}$ M$_{\odot}$. Although this mass limit is more stringent than the Davies limit, our sample is still partly compatible with these predictions.

The work by \citet{2017MNRAS.465.3913B} studied a different class of objects, low-mass haloes (10$^8 <M_{h}/M_{\odot} < 5 \times$ 10$^9$) that have gas in hydrostatic equilibrium with the DM potential and in thermal equilibrium with the UV background radiation (RELHICs). Compared with the dark galaxies by \citet{2006MNRAS.368.1479D}, the RELHICs do not form stars, given the low abundance of \ion{H}{i} and the gas that is in hydrostatic equilibrium, i.e. the gas is stable against gravitational collapse, which could trigger the star formation.  
Compared with the dark galaxies predicted by \citet{2006MNRAS.368.1479D}, their linewidths are dominated by thermal broadening rather than rotation, on the order of a few km s$^{-1}$. The vast majority of the RELHICs show an \ion{H}{i} mass less than 10$^4$ M$_{\odot}$, which is currently below the detection limit of the most sensitive \ion{H}{i} surveys, but the high-mass end of the distribution extends up to 10$^{6.5}$ M$_{\odot}$, so the most massive ones should be in principle detectable. Putting these predictions in context with our sample, we note that none of our candidates is compatible with the RELHICs. However, since the \ion{H}{i} mass limit for the RELHICs is very low, the search for these objects is, for now, limited only to the nearest outskirts of the Milky Way.

In conclusion, it is possible that our DGC sample is formed by two different classes of objects. One is represented by `true' dark galaxies with no stars, that occupy the lower left corner of the W$_{50}$--M$_{\rm HI}$ plane, and `almost' dark galaxies that are simply galaxies with a stellar body so reduced that it is optically dark in the surveys we used.

\subsection{Distances}\label{subsec:dist_peak}
\begin{figure}
    \includegraphics[width=\columnwidth]{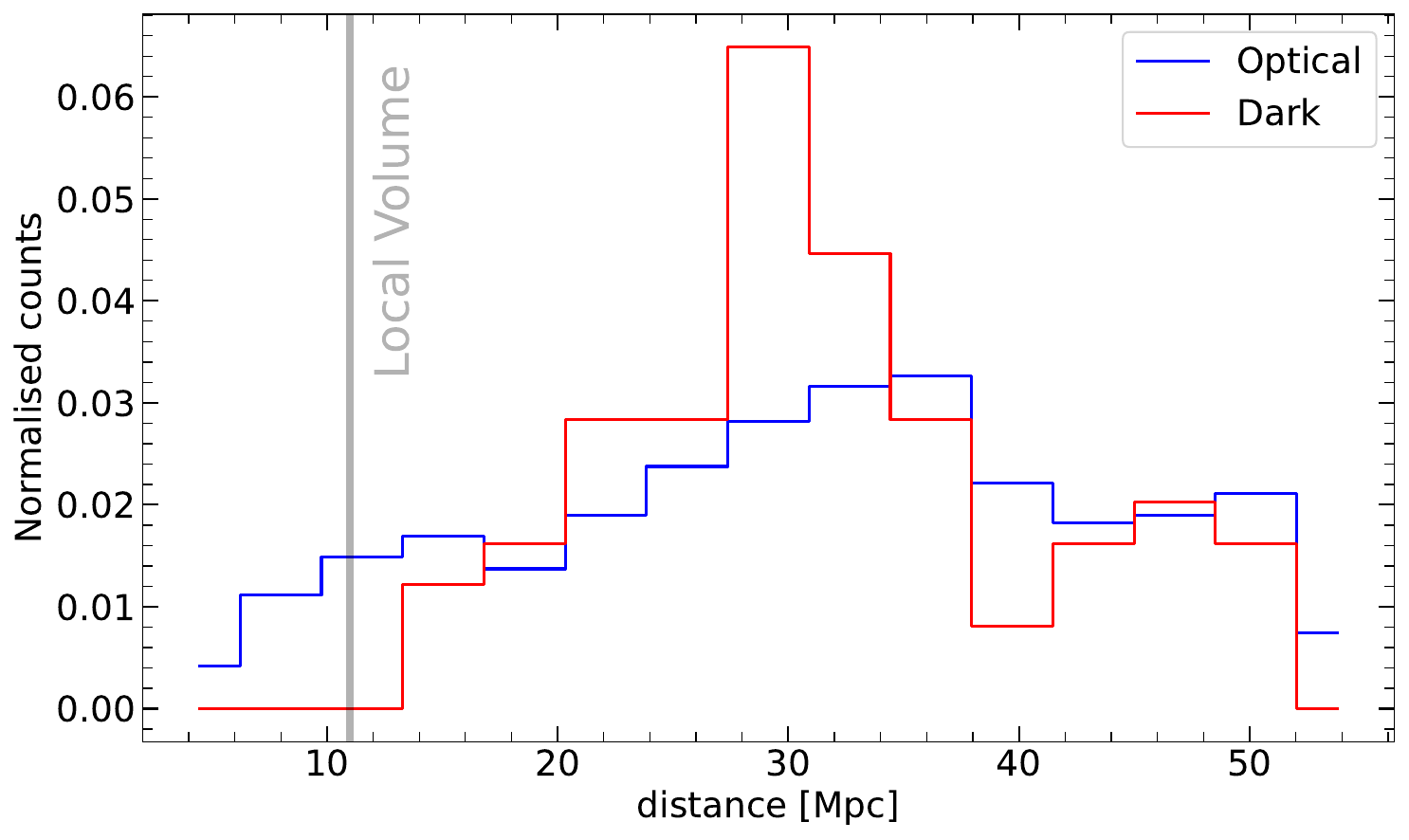}
    \caption{The distribution of the distances for DGCs and for galaxies with an optical counterpart. While the KS test does not indicate any statistically significant difference, it is worth noting that beyond 36 Mpc the number of DGCs drops. None of our DGCs was found within the Local Volume (within 11 Mpc).}
    \label{fig:distance_peak}
\end{figure}

In Figure~\ref{fig:distance_peak}, we show the distribution of distances for both DGCs and galaxies with an optical counterpart. 
Although the KS test for the two distributions does not reveal any statistically significant difference, it is worth discussing some key points. Firstly, we note that all \ion{H}{i} sources with distances less than 14.4 Mpc were successfully cross-matched with the optical catalogues and/or the NED database, or they have an optical counterpart as per Section~\ref{sec:selection_criteria}.
The galaxies with an optical counterpart are evenly distributed across the distance span, while the DGCs show a more peaked distribution, with lower counts beyond 36 Mpc. The origin of this bias is currently unclear, since the sensitivity of the FAST telescope is sufficient to detect DGCs out to 50 Mpc.

We note that the nearest DGC is located at a distance of about 14.4 Mpc, meaning that no DGCs are found within the Local Volume (i.e. within 11 Mpc). Considering that the FASHI survey covers approximately 17.6\% of the entire sky and we identified 70 DGCs, we obtain a volume density of $\rho_{\rm DGC} \simeq 7.6 \times 10^{-4}$ DGC Mpc$^{-3}$. Based on this density, the expected number of DGCs within 11 Mpc over the same sky coverage is $<1$. Considering the entire Local Volume instead, we would expect $\sim 4$ DGCs. With future FASHI releases, we therefore anticipate finding some DGCs within the Local Volume.

Recently, \citet{2026arXiv260104024G} estimated that about eight dark galaxies could be detected by FAST within 2.5 Mpc of the Milky Way, taking into account the telescope’s sensitivity and sky coverage. This result is in tension with our findings, since based on the density derived above, we would expect a number of DGCs within 2.5 Mpc to be $\ll$ 1. However, in the same FASHI catalogue no sources were identified within 2.5 Mpc, with the nearest one located at 2.7 Mpc (UGC 7559).

It should be emphasised that searching for dark \ion{H}{i} sources at such close distances is challenging, as they are not in the Hubble Flow and therefore their radial velocities are dominated by peculiar motions. At these low velocities, they can be confused with compact \citep[][]{1999A&A...341..437B} and ultra-compact \citep[][]{2013ApJ...768...77A} high-velocity clouds, as their \ion{H}{i} profiles are similar to those of DGCs. In a few cases, an overlap along the line-of-sight of multiple clumps could lead to a double-peak emission that could be confused as an indication of ordered rotation \citep[][]{2000A&A...354..853B}. However, these high-velocity clouds have $|cz_{\odot \ \rm radio}|\lesssim 450 \ \mathrm{km \ s^{-1}}$, whereas all DGCs in this work have $cz_{\odot,\rm radio} > 800\ \mathrm{km \ s^{-1}}$, effectively ruling out a local origin.

\section{Results and Conclusions}\label{sec:conclusions}
In this work, we have constructed a catalogue of dark galaxy candidates through a multi-step selection process involving both automated and manual procedures (Section \ref{sec:selection_criteria}). Out of 41741 FASHI \ion{H}{i} sources, we selected the 3436 sources with an \ion{H}{i} recessional velocity lower than 3500 km s$^{-1}$ ($\sim 50$ Mpc). We then performed cross-checks with major optical catalogues of known galaxies, namely the 50MGC, the SGA, and SMUDGes — the latter being used to exclude as many low surface brightness galaxies as possible. After this step, 1278 \ion{H}{i} sources remained without a listed optical counterpart in any of these catalogues.

We then carried out a final check using the NED database, searching for catalogued galaxies located within the 1\farcm5  ~(the radius of the FAST beam) of the \ion{H}{i} source and with a recessional velocity consistent with the \ion{H}{i} detection. Following this step, 1091 sources remained uncatalogued. We then visually inspected all of these sources using deep optical images from the Legacy, and found that out of the 1091 sources 532 (49\%) have a potential optical counterpart; 55 (5\%) lack an optical counterpart but are located within 150 kpc of bright galaxies (and are therefore possible \ion{H}{i} tails or clouds); 419 (38\%) could not be reliably classified and we put in an `unsure' class; 85 (8\%) remain without an unambiguously identified optical counterpart, not near a bright galaxy nor in the unsure class, and are classified as DGCs. We then visually re-checked the DGCs, stacking the $g$, $r$, $i$ images from the Legacy, enhancing the dynamic range of the image, and trying to smooth the images to various scales.

For 14 galaxies with an optical counterpart, we obtained their optical recessional velocities using the Keck telescope. For 13 of them, we verified that their optical velocity is consistent with the radio velocity listed in FASHI, confirming that they are indeed the optical counterparts of the HI sources. The 14$^{\rm th}$ was shown to be a background galaxy, suggesting a small fraction of the optical counterparts may be misidentified.    

After compiling our initial catalogue of 85 DGCs, we proceeded with its final validation by rechecking the sky positions and other parameters listed in the FASHI catalogue. We identified several possible issues within the FASHI catalogue, including possible errors in the recorded positions and recessional velocities (Section \ref{subsec:OffsetPos}); suspiciously clustered sources (a grouping of several DGCs described in Section \ref{subsec:dgccluster}); and sources with very low W$_{50}$ values which, although consistent with theoretical predictions, correspond to spectroscopically unresolved \ion{H}{i} lines in the final FASHI catalogue and may therefore represent residual RFIs (see Section \ref{subsec:lowW50}).  All of these 15 cases have been excluded from the final DGC catalogue, but it is possible that others remain (i.e., those with unusually high W$_{50}$). Although we questioned the reality of them, we do note that catalogued optical galaxy counterparts were identified for 2158 (63\%) out of 3436 FAST \ion{H}{i} sources, and considering also the `O' class the identified sources are 2690 (78\%). Thus, the bulk of the FASHI catalogue is bona fide galaxies. The detection rate for the DGCs is about 2\% (70 sources), and for the uncatalogued optical galaxies is 15.5\% (532 sources), out of 3436 sources.

After the final evaluation, we studied selected properties of the 70 DGCs, comparing them with those of the optical galaxies. The key results are summarised here:
\begin{itemize}    
    \item The first release of the FASHI catalogue covers approximately 17.6\% of the sky, corresponding to a detection rate of about one DGC per 103 deg$^2$. Consequently, once the full FASHI catalogue is completed, we expect roughly 200 DGCs over the area observable by the FAST telescope. The distribution of current DGCs over the sky does not visually show any particular difference from that of optical galaxies, but given the limited sky coverage of the FASHI catalogue and the small number of DGCs, the 2D KS test performed on the spatial distributions is not conclusive at this stage. The full release of the FASHI catalogue will help to address this point.

    \item We analysed the distribution of DGCs and galaxies with an optical counterpart in the W$_{50}$ -- M$_{\mathrm{HI}}$ plane, and the DGCs appear to have a different distribution, confirmed by a KS test. As expected, they contain, on average, lower hydrogen mass than the galaxies with an optical counterpart, but still have a W$_{50}$ distribution compatible with that of the optical galaxies. However, a few of the DGCs exhibit an \ion{H}{i} mass $\gtrsim 10^{8.5}$ M$_{\odot}$ and $W_{50}\gtrsim 180$ km s$^{-1}$, but without an optical counterpart. We speculate that these sources could be formed by a superposition of different \ion{H}{i} clouds along the line of sight.

    \item We compared our DGC sample with the dark galaxy candidates from ALFALFA found by \citep[][K25]{2025ApJS..279...38K}, and candidates from WALLABY Pilot Surveys by \citep[][OB25]{2025PASA...42...87O} finding that our DGCs show a lower \ion{H}{i} mass at fixed W$_{50}$.

    \item We studied the distribution of our DGC sample with the theoretical predictions by \citealt{2006MNRAS.368.1479D} (rotating dark galaxies) and by \citealt{2017MNRAS.465.3913B} (RELHICs), finding that most of our DGCs are compatible with the predictions of \citet{2006MNRAS.368.1479D}, but none of them are compatible with the RELHICs. Since both theoretical predictions are about systems with no stars at all, it is possible that our DGC sample is formed by two different classes of objects: one, composed of dark galaxies without a stellar body (the DGCs that are compatible with the \citealt{2006MNRAS.368.1479D} predictions), and another composed of `almost' dark galaxies with a stellar body below the detection limit of the optical surveys we used.

    \item We analysed the distance distribution of DGCs and optical galaxies, and although there is no statistically significant difference between the two distributions, the DGCs show lower counts beyond 36 Mpc. The origin of this bias is currently unclear, since, given the average peak flux of the DGCs, they could in principle be detected with an S/N $>5$ out to distances of up to 50 Mpc. The closest DGC has a distance of 14.4 Mpc (i.e. none were found in the Local Volume).
      
\end{itemize}

We conclude this work with a few suggestions for future research. Even though our DGCs are the most promising candidates among the FASHI sources for being true dark galaxies, they still need to be confirmed through independent \ion{H}{i} observations with both high spectral and spatial resolution and followed up by deep optical imaging to search for a faint optical counterpart. In this way, it may be possible to determine an \ion{H}{i} size and inclination, and total stellar mass (or upper limit). For the latter, telescopes such as the VST \citep[][]{2024A&A...689A.306S}  or the new generation of optical telescopes, like the Vera Rubin Observatory and Euclid, will provide valuable insight. Moreover, citizen science projects could also be envisioned, involving amateur astronomers, who have already demonstrated the ability to obtain extremely deep images \citep[][]{2025arXiv250402071M}.

\section*{Acknowledgements}
{\small We thank the anonymous referee for the insightful comments that helped us improve the original manuscript.}
{\small MM warmly thanks Chuan-Peng Zhang and Chen Xu for the extremely useful discussion about possible issues in the first release of the FASHI catalogue and for providing the \ion{H}{i} spectra.}
{\small MM thanks Arianna Di Cintio and Guacimara Garc\'ia-Bethencourt for useful discussions.}
{\small This work has used the data from the Five-hundred-meter Aperture Spherical radio Telescope (FAST). FAST is a Chinese national mega-science facility, operated by the National Astronomical Observatories of Chinese Academy of Sciences (NAOC).}
{\small Some of the data presented herein were obtained at Keck Observatory, which is a private non-profit organisation operated as a scientific partnership among the California Institute of Technology, the University of California, and the National Aeronautics and Space Administration. The Observatory was made possible by the generous financial support of the W. M. Keck Foundation. The authors also wish to recognise and acknowledge the very significant cultural role and reverence that the summit of Maunakea has always had within the Native Hawaiian community. We are most fortunate to have the opportunity to conduct observations from this mountain.}
{\small This research has made use of the Keck Observatory Archive (KOA), which is operated by the W. M. Keck Observatory and the NASA Exoplanet Science Institute (NExScI), under contract with the National Aeronautics and Space Administration.}
{\small MM acknowledges financial support received through a Swinburne University Postgraduate Research Award throughout the making of this work.}
{\small We thank the Australian Research Council (ARC) for the financial support via the Discovery Project DP250101673.}
{\small This research has made use of the NASA/IPAC Extragalactic Database (NED),
which is operated by the Jet Propulsion Laboratory, California Institute of Technology,
under contract with the National Aeronautics and Space Administration.}
{\small This research has made use of the VizieR catalogue access tool, CDS, Strasbourg, France.}
{\small The Parkes telescope is part of the Australia Telescope which is funded by the Commonwealth of Australia for operation as a National Facility managed by CSIRO.}
{\small The Siena Galaxy Atlas was made possible by funding support from the U.S. Department of Energy, Office of Science, Office of High Energy Physics under Award Number DE-SC0020086 and from the National Science Foundation under grant AST-1616414.}
{\small We acknowledge the usage of the HyperLeda database (http://leda.univ-lyon1.fr).}
{\small The Legacy Surveys consist of three individual and complementary projects: the Dark Energy Camera Legacy Survey (DECaLS; Proposal ID \#2014B-0404; PIs: David Schlegel and Arjun Dey), the Beijing-Arizona Sky Survey (BASS; NOAO Prop. ID \#2015A-0801; PIs: Zhou Xu and Xiaohui Fan), and the Mayall z-band Legacy Survey (MzLS; Prop. ID \#2016A-0453; PI: Arjun Dey). DECaLS, BASS and MzLS together include data obtained, respectively, at the Blanco telescope, Cerro Tololo Inter-American Observatory, NSF’s NOIRLab; the Bok telescope, Steward Observatory, University of Arizona; and the Mayall telescope, Kitt Peak National Observatory, NOIRLab. Pipeline processing and analyses of the data were supported by NOIRLab and the Lawrence Berkeley National Laboratory (LBNL). The Legacy Surveys project is honored to be permitted to conduct astronomical research on Iolkam Du’ag (Kitt Peak), a mountain with particular significance to the Tohono O’odham Nation.
NOIRLab is operated by the Association of Universities for Research in Astronomy (AURA) under a cooperative agreement with the National Science Foundation. LBNL is managed by the Regents of the University of California under contract to the U.S. Department of Energy.
This project used data obtained with the Dark Energy Camera (DECam), which was constructed by the Dark Energy Survey (DES) collaboration. Funding for the DES Projects has been provided by the U.S. Department of Energy, the U.S. National Science Foundation, the Ministry of Science and Education of Spain, the Science and Technology Facilities Council of the United Kingdom, the Higher Education Funding Council for England, the National Center for Supercomputing Applications at the University of Illinois at Urbana-Champaign, the Kavli Institute of Cosmological Physics at the University of Chicago, Center for Cosmology and Astro-Particle Physics at the Ohio State University, the Mitchell Institute for Fundamental Physics and Astronomy at Texas A\&M University, Financiadora de Estudos e Projetos, Fundacao Carlos Chagas Filho de Amparo, Financiadora de Estudos e Projetos, Fundacao Carlos Chagas Filho de Amparo a Pesquisa do Estado do Rio de Janeiro, Conselho Nacional de Desenvolvimento Cientifico e Tecnologico and the Ministerio da Ciencia, Tecnologia e Inovacao, the Deutsche Forschungsgemeinschaft and the Collaborating Institutions in the Dark Energy Survey. The Collaborating Institutions are Argonne National Laboratory, the University of California at Santa Cruz, the University of Cambridge, Centro de Investigaciones Energeticas, Medioambientales y Tecnologicas-Madrid, the University of Chicago, University College London, the DES-Brazil Consortium, the University of Edinburgh, the Eidgenossische Technische Hochschule (ETH) Zurich, Fermi National Accelerator Laboratory, the University of Illinois at Urbana-Champaign, the Institut de Ciencies de l’Espai (IEEC/CSIC), the Institut de Fisica d’Altes Energies, Lawrence Berkeley National Laboratory, the Ludwig Maximilians Universitat Munchen and the associated Excellence Cluster Universe, the University of Michigan, NSF’s NOIRLab, the University of Nottingham, the Ohio State University, the University of Pennsylvania, the University of Portsmouth, SLAC National Accelerator Laboratory, Stanford University, the University of Sussex, and Texas A\&M University.
BASS is a key project of the Telescope Access Program (TAP), which has been funded by the National Astronomical Observatories of China, the Chinese Academy of Sciences (the Strategic Priority Research Program “The Emergence of Cosmological Structures” Grant \# XDB09000000), and the Special Fund for Astronomy from the Ministry of Finance. The BASS is also supported by the External Cooperation Program of Chinese Academy of Sciences (Grant \# 114A11KYSB20160057), and Chinese National Natural Science Foundation (Grant \# 12120101003, \# 11433005).
The Legacy Survey team makes use of data products from the Near-Earth Object Wide-field Infrared Survey Explorer (NEOWISE), which is a project of the Jet Propulsion Laboratory/California Institute of Technology. NEOWISE is funded by the National Aeronautics and Space Administration.
The Legacy Surveys imaging of the DESI footprint is supported by the Director, Office of Science, Office of High Energy Physics of the U.S. Department of Energy under Contract No. DE-AC02-05CH1123, by the National Energy Research Scientific Computing Center, a DOE Office of Science User Facility under the same contract; and by the U.S. National Science Foundation, Division of Astronomical Sciences under Contract No. AST-0950945 to NOAO.}

\section*{Data Availability}
This study is based almost entirely on publicly available data from the FASHI, SGA, 50MGC, SMUDGes, and NED catalogues. For the FASHI spectra, please refer to \citet{2024SCPMA..6719511Z}. The Keck Observatory data will be made publicly available via the Keck Observatory Archive (KOA) after the standard proprietary period of 12 months.



\bibliographystyle{mnras}
\bibliography{biblio} 


\appendix

\section{Keck spectroscopic confirmation of selected optical counterparts}\label{app:KeckSpectra}

We selected the optical counterparts through a visual assessment that considered both morphological and positional aspects. We find that the vast majority of successfully cross-matched galaxies with optical catalogues are blue, star-forming systems located near the centre of the FAST beam.
However, we stress that, since the confirmation of previously uncatalogued optical galaxies associated with the \ion{H}{i} detections is not the primary scope of this work, no quantitative criteria were applied in the selection of potential optical counterparts. 

Fourteen sources from the optical counterpart list were selected for spectroscopic confirmation during Keck observing runs on 2025 April 21st, 22nd, and 30th.
Since the spectroscopic confirmation of these objects was not the main science goal of these nights, we selected bright targets to be observed as `fillers' during twilight using short exposures ($\sim$5–10 minutes). 
The observations were performed using the Keck Cosmic Web Imager integral field spectrograph \citep[KCWI;][]{2018ApJ...864...93M, 2024SPIE13096E..47M}, mounted on the Keck II Nasmyth platform. KCWI allows for different configurations on the blue and red sides, covering the wavelength ranges 3500--5600~\AA~ and 5400--10800~\AA, respectively. The selected configurations were: BL grating (low resolution, blue side) centred at 4550 \AA, with the medium slicer, providing a field of view of 16\farcs5$\times$20\farcs4; RH2 (high resolution, red side) centred at 6750 \AA \ with the medium slicer. Since our goal was to measure radial velocities, we used only the blue spectra, which are much less affected by cosmic rays compared to the red side.
For the data reduction, we followed the procedure described by \citet{levitskiy2025}, and the full-spectrum fit was performed using the \texttt{pPXF} software package \citep[][]{2004PASP..116..138C, Cappellari2023}

We have noticed that several spectra exhibit both absorption and emission lines, the latter of which are associated with star-forming regions. Rather than masking the emission features, we adopted a spectral template composed of two components: one containing only absorption lines from the stellar population, and another including emission lines from the gas (Balmer lines plus forbidden lines). The procedure we followed is essentially the same as the one illustrated in one of the \texttt{pPXF} usage examples (\texttt{ppxf\_example\_population\_gas\_sdss.ipynb}\footnote{\url{https://github.com/micappe/ppxf_examples/blob/main/ppxf_example_population_gas_sdss.ipynb}}). 
We allowed the systemic velocity for each gas and stellar template to be different, and we used the $cz_{\odot,\mathrm{radio}}$ as the initial guess for the recessional velocity for each galaxy. 

\begin{table}
    \centering
    \caption{Results of the pPXF fitting on optical spectra for the selected sample. In the first column, the FASHI name of the source is reported; in the second column, the heliocentric velocity as listed in the FASHI catalogue is reported; and in the third column, the stellar velocity inferred from the \texttt{pPXF} fitting of the KCWI spectra is reported. The galaxy we selected as the potential optical counterpart of the source J150625.78+452302.6 (an asterisk highlights the optical stellar velocity) turned out to be a background galaxy unrelated to the FASHI source. More details in the text. 
    The sources are listed in the same order as in Figure \ref{fig:montageKeck}.}
    \label{tab:velocityKeck}
    \begin{tabular}{lll}
        \hline
        \textbf{Name} & $\mathbf{cz_{\odot,\mathrm{radio}}}$ \textbf{[km s$^{-1}$]} & $\mathbf{cz_{\odot,\mathrm{star}}}$ \textbf{[km s$^{-1}$]} \\
        \hline
        J165132.43+581535.3 & $3257.23 \pm 1.75$ & $3176 \pm 42$ \\
        J164710.53+542531.0 & $2107.34 \pm 1.01$ & $2075 \pm 22$ \\
        J162433.80+593318.5 & $2849.40 \pm 0.73$ & $2853 \pm 16$ \\
        J161544.84+595424.3 & $3003.18 \pm 1.57$ & $2978 \pm 35$ \\
        J161525.11-015011.8 & $1546.79 \pm 1.47$ & $1570 \pm 23$ \\
        J155532.07+623616.4 & $825.37 \pm 0.73$ & $850 \pm 30$ \\
        J153301.92+570153.0 & $3101.57 \pm 1.92$ & $3098 \pm 29$ \\
        J152447.04+550732.6 & $2621.8 \pm 1.34$ & $2633 \pm 41$ \\
        J152242.17+623424.6 & $494.98 \pm 0.78$ & $538 \pm 26$ \\
        J150625.78+452302.6 & $1089.4 \pm 1.41$ & $16337 \pm 35$ $(\ast)$ \\
        J132302.81+390949.5 & $2415.10 \pm 0.56$ & $2393 \pm 24$ \\
        J092633.35+481347.1 & $662.85 \pm 1.44$ & $674 \pm 26$ \\
        J090231.75+621235.7 & $793.93 \pm 0.98$ & $800 \pm 46$ \\
        J080428.28+405843.1 & $2534.41 \pm 2.01$ & $2524 \pm 38$ \\
        \hline
    \end{tabular}
\end{table}

\begin{figure*}
    \includegraphics[width=0.9\textwidth]{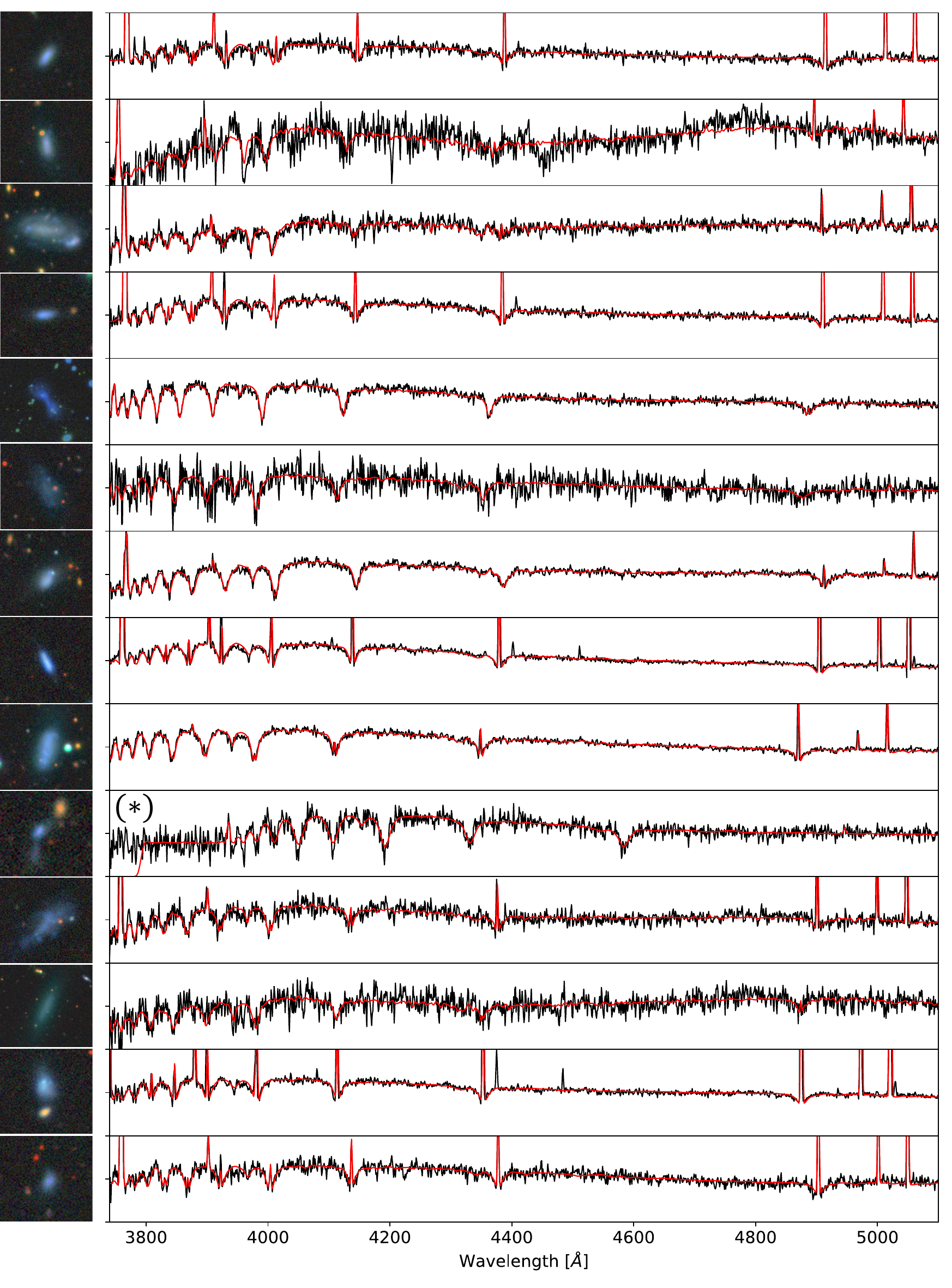}
    \caption{A montage of 14 optical counterpart spectra obtained with KCWI at the Keck telescope. The galaxies are shown from the top in the same order as in Table \ref{tab:velocityKeck}, which reports the \ion{H}{i} velocities and the optical velocities. For each galaxy, a small cutout from the Legacy is shown on the left. The size of the cutouts varies slightly from one galaxy to another to better highlight the galaxy itself. On the right, the integrated galaxy spectrum (black line) and the fit obtained through pPXF (red line) are shown. Note that several galaxies reveal emission lines. The galaxy marked with an asterisk (J150625.78+452302.6) has a redshift significantly higher than that of the other sources, as it is a background galaxy not associated with the \ion{H}{i} source identified by FAST. See the text for further details.}
    \label{fig:montageKeck}
\end{figure*}

Figure \ref{fig:montageKeck} displays a montage of the optical spectra, with a small cutout of each galaxy taken from the Legacy. On the right side of each cutout, the galaxy spectrum (black line) and the best-fit spectrum given by \texttt{pPXF} (red line, where the stellar component and the gas component were combined to form the total template) are shown.

Table \ref{tab:velocityKeck} summarises the results of the \texttt{pPXF} fitting routine. The first column reports the source name as listed in the FASHI catalogue, the second column reports the $cz_{\odot,\mathrm{radio}}$ as listed in the FASHI catalogue, and the third column lists the stellar velocities from the optical spectra.

\begin{figure}
    \includegraphics[width=\columnwidth]{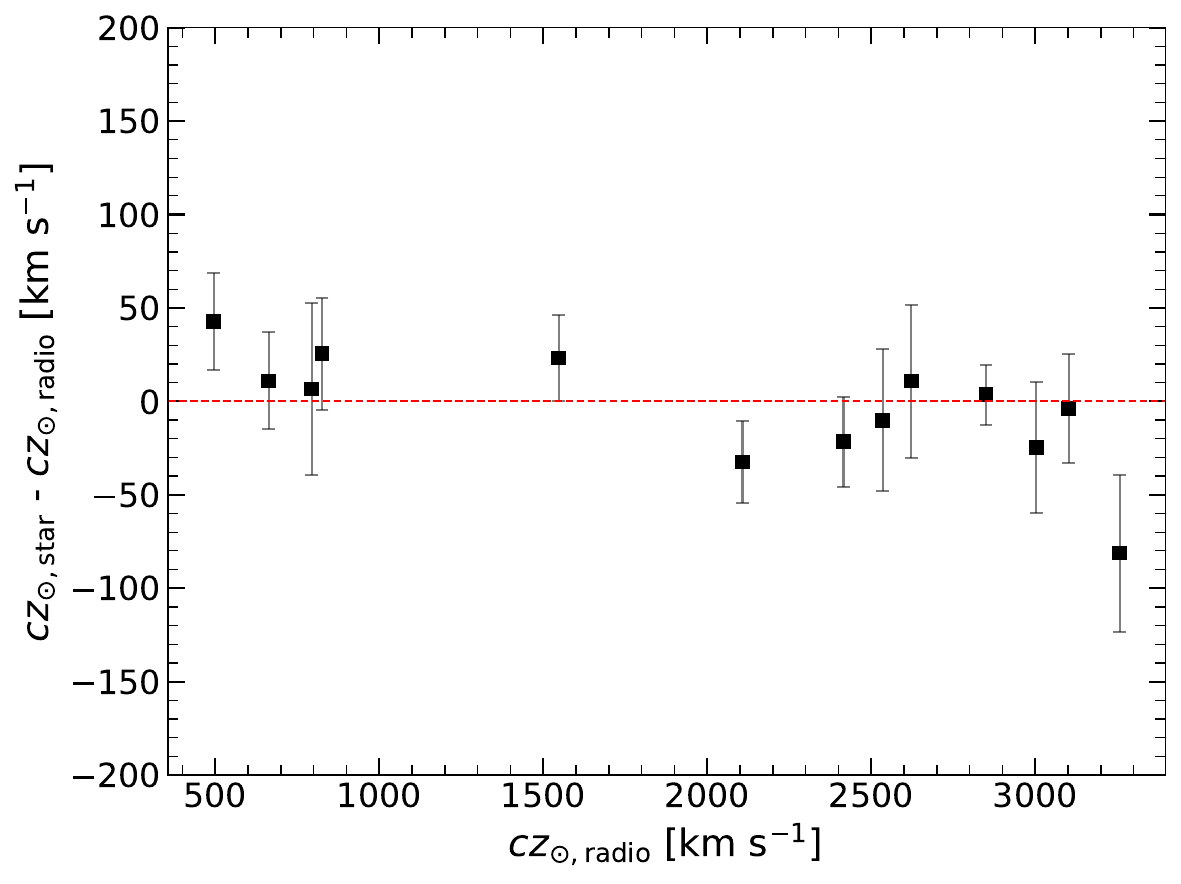}
    \caption{The relation between the optical velocities inferred from the Keck spectra and the FAST \ion{H}{i} velocity ($cz_{\odot,\mathrm{radio}}$). All optical radial velocities are consistent with \ion{H}{i} velocities within 2$\sigma$.}
    \label{fig:velocity_offsets}
\end{figure}

Figure \ref{fig:velocity_offsets} shows the relation between the $cz_{\odot, \mathrm{radio}}$ and the stellar velocities obtained. Out of 14 sources, 13 were confirmed to be indeed the optical counterparts of the \ion{H}{i} detections, since all optically-inferred radial velocities are compatible within 2$\sigma$ with \ion{H}{i} velocities.

In only one case (J150625.78+452302.6, which is not shown in Figure \ref{fig:velocity_offsets}), the velocity of the potential optical counterpart did not match the \ion{H}{i} velocity. In this case, we first estimated the redshift by examining the observed wavelengths of the Ca H and K lines, and then we reran \texttt{pPXF}. The result is highlighted by an asterisk next to the \texttt{pPXF}-inferred velocity in Table \ref{tab:velocityKeck}, and the relative spectrum is marked by an asterisk in Figure \ref{fig:montageKeck}. 

\begin{figure}
    \includegraphics[width=\columnwidth]{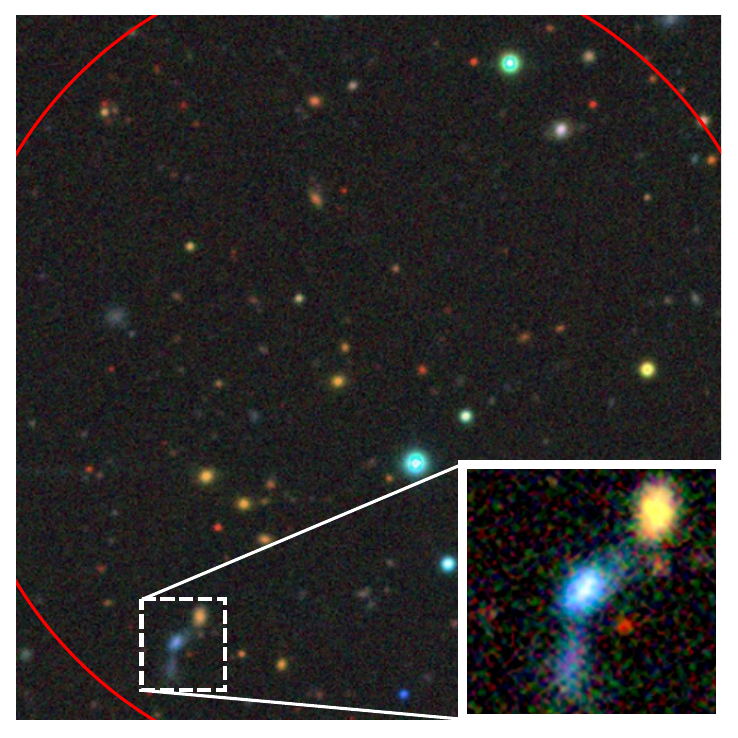}
    \caption{A cutout from the Legacy of the FASHI source J150625.78+452302.6. The incomplete red circle highlights the 3\arcmin ~FAST beam, with the \ion{H}{i} source located at the centre of the cutout, and the inset shows a stretched and zoomed detail of the galaxy. This galaxy was selected as the potential optical counterpart and was observed at Keck \ref{app:KeckSpectra}, revealing that it is a background galaxy.}
    \label{fig:J150625_case}
\end{figure}

Figure \ref{fig:J150625_case} illustrates this case. In the absence of other sources that are well compatible with the nature of the other optical counterparts (bright and near the beam centre), we selected the source highlighted by the dashed white rectangle, although the positional matching with the \ion{H}{i} source is rather poor; however, as previously indicated, this source turned out to be a background galaxy.
The insert displays two other galaxies: one, orange, in the upper-right corner and another, just below the bright blue galaxy in the centre. Since they are both within KCWI field, we tried to recover their recessional velocities as well. The orange galaxy turned out to be a background galaxy with $cz_{\rm \odot, star}\sim 87400$ km s$^{-1}$. The galaxy below the central blue one is too faint to obtain a recessional velocity.

We decided to classify this source as `unsure', as it is located in a fairly crowded field with other potential optical counterparts, and the most plausible explanation is a misidentification of the optical counterpart. This case highlights the importance of optical redshifts for identifying optical counterparts and may warrant further follow-up.

\section{The search for optical counterparts}\label{app:optCountQuest}
In this Appendix, we discuss and comment on the search for optical counterparts using \ion{H}{i} detections, discussing both cases we found in the FASHI catalogue \citep{2024SCPMA..6719511Z} and in the work by \citet{2025ApJS..279...38K}. We also briefly comment on the recent findings regarding the `dark' galaxy first identified by \citet{2023ApJ...944L..40X}.

\citet{2024SCPMA..6719511Z} performed a cross-match on the FASHI catalogue using the SGA, the SDSS spectroscopic, and SDSS photometric catalogues. 
Since our selection process was completely unbiased with respect to the cross-matching performed by \citet{2024SCPMA..6719511Z}, it is possible that some DGCs are included in those cross-matches. Obviously, none of our DGCs was found among the SDSS spectroscopic counterparts identified in the FASHI catalogue \citep[table 5 in ][]{2024SCPMA..6719511Z}. On the other hand, several of our DGCs were identified in the SDSS photometric counterparts catalogue \citep[table 6 in][]{2024SCPMA..6719511Z}. However, since the cross-matching is solely based on positional coincidence, and the SDSS optical counterparts are generally faint or lie in the background (otherwise they would have been identified during our visual inspection process), these matches should be considered as spurious. We discuss here two illustrative examples:
\begin{itemize}
\item The source J150045.42+332746.9 was identified by us as a reliable DGC, but was classified as having a possible optical counterpart found only in the SDSS photometric catalogue: a galaxy located approximately 100\arcsec \ from the centre of the beam. We found that this optical source has a DESI spectrum and has been confirmed as a galaxy with a $cz_{\odot} \simeq 31780$ km s$^{-1}$, which is significantly different from the FASHI $cz_{\odot,\mathrm{radio}} \simeq 1852$ km s$^{-1}$. Therefore, this SDSS galaxy is not associated with the \ion{H}{i} source and represents a spurious match.

\item The source J124838.01+500346.8 ($cz_{\odot,\mathrm{radio}} = 2230.9 \pm 2.14$ km s$^{-1}$) is located 50\arcsec \ north-west of a galaxy listed in the Siena Galaxy Atlas  (2MASXJ12484339+5003287, $cz_{\odot} = 31507.2 \pm 13.1$ km s$^{-1}$), which was considered as the photometric optical counterpart. On the other hand, we validated the \ion{H}{i} source as a good DGC, since 2MASXJ12484339+5003287 has too high recessional velocity, and therefore was filtered out by our first velocity threshold of 3500 km s$^{-1}$. Furthermore, given the frequency range of FASHI spectra, which corresponds to a velocity up to $\sim 26400$ km s$^{-1}$, this 2MASS galaxy cannot have been detected.
\end{itemize}

These cases show the importance of matching \ion{H}{i} and optical velocities when identifying counterparts. This is also evident in our spectroscopic confirmation, as one galaxy was found to be a background source.

Although the procedure followed by \citet{2025ApJS..279...38K} for selecting dark galaxy candidates from ALFALFA is very similar to ours, the visual inspection aimed at identifying possible optical counterparts may have led to different samples. For example, the source AGC 208881, considered a valid dark galaxy candidate by \citet{2025ApJS..279...38K}, shows a (very faint) optical counterpart in the Legacy; in this case, we would have categorised this source as an \ion{H}{i} detection with an optical counterpart (class `O'). The source AGC 208431 has a small blue galaxy located approximately 83\arcsec ~from the position of the dark galaxy candidate, thereby within the Arecibo beam, although slightly off-axis. No optical redshift is available for this galaxy, so it is not possible to establish whether it is indeed the optical counterpart of the \ion{H}{i} emission. We also note that some dark galaxy candidates from \citet{2025ApJS..279...38K} lack optical images in Legacy, making the visual inspection challenging, as shallower optical surveys must be used instead. 

The study of J0139+4328 provides an additional point for reflection. \citet{2023ApJ...944L..40X} first indicated this object as a dark galaxy candidate, since they were unable to identify any optical counterpart in Pan-STARRS1 images, given the lack of accurate positional information. Later, \citet{2026A&A...708A..40S}, using the VLA, obtained a better centroiding accuracy for the \ion{H}{i} source, which led to the discovery of a faint optical counterpart visible in Pan-STARRS1 stacked image. \citet{2026A&A...705L...9M} confirmed through deep imaging and spectroscopy that this newly discovered galaxy is indeed the optical counterpart of the \ion{H}{i} detection. This highlights how the identification of DGCs is extremely challenging and, to be carried out robustly, requires single-dish \ion{H}{i} wide surveys, spatially resolved \ion{H}{i} data, and extremely deep optical imaging.

All considered, we wish to emphasise that the selection of the optical counterparts, although performed with care, is carried out manually, taking into account several factors such as the colour, the photometric redshift, the angular size relative to what is expected from the \ion{H}{i} disc extent, and the spatial matching with the HI detection position. Alternative approaches may lead to differences in the assignment of optical counterparts.

\section{Sources with incorrect listed position and velocity}\label{app:posIssues}
In this Appendix, we describe the five excluded detections at declination $=-5$\degr, in which the FAST telescope detected an \ion{H}{i} emission without an optical counterpart but suspiciously near luminous galaxies, which were not detected by FAST, but were detected by HIPASS or previous \ion{H}{i} studies. All the cases discussed so far share some similarities: first, the offset is almost entirely in right ascension; second, they are all located in the same strip at about Dec = -5\degr; third, the difference between the radial velocities measured by FAST and those tabulated in the previous studies for the near bright galaxy is very similar in all cases, about 2000 km s$^{-1}$, which at 1420 MHz corresponds to a frequency shift of about 9.5 MHz. In Table \ref{tab:fastIssues} we summarise the \ion{H}{i} properties of the five FAST detections and the luminous nearby galaxies.

\begin{figure}
    \includegraphics[width=\columnwidth]{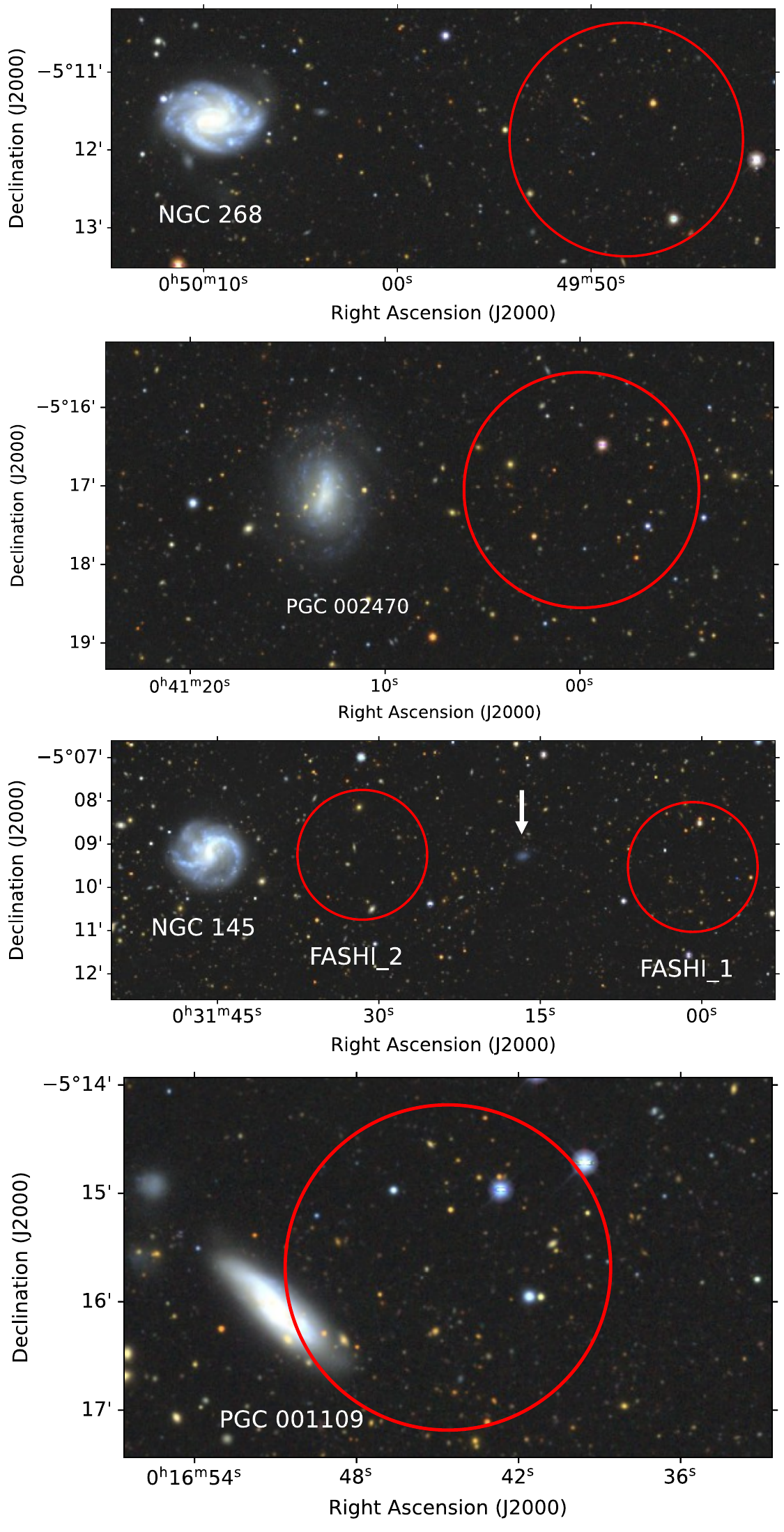}
    \caption{Position offset cases of NGC 268, PGC 002470, NGC 145, and PGC 001109. In all four panels, the red circle indicates the FAST beam size of 3\arcmin ~in diameter, centred at the position listed in the FASHI catalogue. In each case, the luminous labelled galaxy was not detected by FAST, but was detected by HIPASS or previous \ion{H}{i} studies. The beam size of HIPASS is approximately 15\arcmin; therefore, it is not shown in the images. The white arrow in the NGC 145 panel indicates the optical counterpart that is probably linked with the FASHI\_1 detection. It is worthwhile noting that the position offset is almost exclusively in right ascension for all the sources, and they are all located at declination $= -5\degr$.}
    \label{fig:issues_catalogue_positionOffset}
\end{figure}

 \begin{table*}
 \caption{Properties of selected sources detected in FASHI and of the nearby galaxy. The HIPASS catalogue doesn't list the errors for the quantities listed here. 
 For each source, we list the position (RA and Dec), the heliocentric velocity ($cz_{\odot}$), the velocity width of the line at 50\% and 20\% level of the peak flux density (W$_{50}$ and W$_{20}$), the peak flux density (F$_{\mathrm{P}}$), and the integrated flux density over the \ion{H}{i} line (S$_{\mathrm{sum}}$).}
 \label{tab:fastIssues}
\begin{tabular}{|c||c|c|c|c|c|c|c|}
\hline
\multicolumn{7}{|c|}{\textbf{NGC 268 (FASHI J004948.20-051152.6)}} \\
\hline
   & {\bf RA [deg]} & {\bf DEC [deg]} & {\bf $\mathbf{cz_{\odot}}$ [km s$^{-1}$]} & {\bf W$\mathbf{_{50}}$ [km s$^{-1}$]} & {\bf W$\mathbf{_{20}}$ [km s$^{-1}$]} & {\bf F$\mathbf{_P}$ [mJy]} & {\bf S$\mathbf{_{sum}}$ [Jy \ km \ s$^{-1}$]} \\
{\bf FASHI} & 12.45079 & -5.197778 & 3456.16 $\pm$ 0.42 & 239.67 $\pm$ 0.85 & 262.97 $\pm$ 1.27 & 48.36 $\pm$ 3.7 & 8.542 $\pm$ 0.169 \\
{\bf HIPASS} & 12.53792 & -5.184167 & 5418.8 & 240.2 & 279.5 & 49.7 & 10.9 \\
\hline
\multicolumn{7}{|c|}{\textbf{PGC 002470 (FASHI J004059.93-051703.0)}} \\
\hline
   & {\bf RA [deg]} & {\bf DEC [deg]} & {\bf $\mathbf{cz_{\odot}}$ [km s$^{-1}$]} & {\bf W$\mathbf{_{50}}$ [km s$^{-1}$]} & {\bf W$\mathbf{_{20}}$ [km s$^{-1}$]} & {\bf F$\mathbf{_P}$ [mJy]} & {\bf S$\mathbf{_{sum}}$ [Jy \ km \ s$^{-1}$]} \\
{\bf FASHI} & 10.2497 & -5.2842 & 1553.75 $\pm$ 0.35 & 120.16 $\pm$ 0.69 & 132.04 $\pm$ 1.04 & 53.00 $\pm$ 3.77 & 5.349 $\pm 0.117$\\
{\bf HIPASS} & 10.3198 & -5.2853 & 3584.1 & 124.8 & 195.1 & 46.6 & 6.5\\
\hline
\multicolumn{7}{|c|}{\textbf{NGC 145 (FASHI\_1 J003100.83-050931.6 and FASHI\_2 J003131.53-050914.9)}} \\
\hline
   & {\bf RA [deg]} & {\bf DEC [deg]} & {\bf $\mathbf{cz_{\odot}}$ [km s$^{-1}$]} & {\bf W$\mathbf{_{50}}$ [km s$^{-1}$]} & {\bf W$\mathbf{_{20}}$ [km s$^{-1}$]} & {\bf F$\mathbf{_P}$ [mJy]} & {\bf S$\mathbf{_{sum}}$ [Jy \ km \ s$^{-1}$]} \\
{\bf FASHI\_1} & 7.7535 & -5.1588 & 2117.49 $\pm$ 0.84 & 41.14 $\pm$ 1.68 & 59.73 $\pm$ 2.52 & 18.74 $\pm$ 1.78 & 0.718 $\pm 0.041$\\
{\bf FASHI\_2} & 7.8814 & -5.1541 & 2117.95 $\pm$ 0.24 & 91.3 $\pm$ 0.49 & 145.74 $\pm$ 0.73 & 173.73 $\pm$ 4.5 & 17.383 $\pm 0.225$\\
{\bf HIPASS} & 7.9296 & -5.1719 & 4125.4 & 85.8 & 148.9 & 139.4 & 13.3\\
\hline
\multicolumn{7}{|c|}{\textbf{PGC 001109 (FASHI J001644.63-051541.2)}} \\
\hline
   & {\bf RA [deg]} & {\bf DEC [deg]} & {\bf $\mathbf{cz_{\odot}}$ [km s$^{-1}$]} & {\bf W$\mathbf{_{50}}$ [km s$^{-1}$]} & {\bf W$\mathbf{_{20}}$ [km s$^{-1}$]} & {\bf F$\mathbf{_P}$ [mJy]} & {\bf S$\mathbf{_{sum}}$ [Jy \ km \ s$^{-1}$]} \\
{\bf FASHI} & 4.1859 & -5.2614 & 1924.64 $\pm$ 0.52 & 261.71 $\pm$ 1.04 & 288.87 $\pm$ 1.56 & 29.84 $\pm$ 3.28 & 7.159 $\pm$ 0.164\\
{\bf HIPASS} & - & - & - & - & - & - & -\\
{\bf Theureau et al. (1988)} & 4.2139 & -5.2722 & 3958 $\pm$ 10 & 266 $\pm$ 20 & 303 $\pm$ 30 & - & -\\ 
\hline
\end{tabular}
 \end{table*}

\subsection{The case of NGC 268}\label{subsec:ngc268}
The FASHI source J004948.20-051152.6 does not appear to have a clear optical counterpart, but its position projected on the sky is located near the galaxy NGC 268, about 5\arcmin \ off to the west (see the upper panel in Figure \ref{fig:issues_catalogue_positionOffset}). 
We first classified this source as a good dark galaxy candidate, since the $cz_{\odot,\mathrm{radio}} = 3456.16 \pm 0.42$ km s$^{-1}$ of the \ion{H}{i} source is quite different to the NGC 268 optical radial velocity of 5494 $\pm$ 4 km s$^{-1}$ \citep[][]{2005AJ....130.1037C}, and it is outside of the 3\arcmin \ beam, so the FASHI detection could not be a tidal tail or a debris from NGC 268. However, we noticed that NGC 268 is not listed in the FASHI catalogue, even though it has been identified by HIPASS (which has lower sensitivity). 
The principal properties of the FASHI source and the \ion{H}{i} emission from NGC 268 detected by HIPASS are listed in Table \ref{tab:fastIssues}. Except for $cz_{\odot,\mathrm{radio}}$ and RA and Dec, the W$_{50}$, W$_{20}$, peak and integrated flux density appear comparable. Given that the line properties are virtually identical, we conclude that the FASHI source 004948.20-051152.6 is the \ion{H}{i} emission of NGC 268, but listed in the FASHI catalogue with a position offset and with a wrong radial velocity.

\subsection{The case of PGC 002470}\label{subsec:pgc002470}
The case of PGC 002470 (see the second panel in Figure \ref{fig:issues_catalogue_positionOffset} and Table \ref{tab:fastIssues}) is almost identical to the NGC 268 case. PGC 002470 was detected by HIPASS, but not by FAST. The FASHI catalogue lists a source near PGC 002470, approximately 4\arcmin \ off the galaxy (in this case, the offset is almost entirely in right ascension). Unlike the NGC 268 case, W$_{50}$ are similar and W$_{20}$ are slightly different between FASHI and HIPASS; the systemic velocity inferred by HIPASS is $\sim 2000$ km s$^{-1}$ greater than that measured by FAST. 

\subsection{The case of NGC 145}
The case of NGC 145 is perhaps the most emblematic (third panel in Figure \ref{fig:issues_catalogue_positionOffset}). Indeed, NGC 145 itself was not detected by FAST; however, two detections were found close to this galaxy, whose properties are listed in Table \ref{tab:fastIssues}. Once again, the offset is almost entirely in right ascension, and moreover, the detection closest to NGC 145 (FASHI\_2, J003131.53-050914.9) shows a more prominent \ion{H}{i} line (its peak flux density is about ten times higher than that of the FASHI\_1, J003100.83-050931.6). The $cz_{\rm \odot, radio}$ of the FAST detections are nearly identical. As indicated by the white arrow in the lower panel, there is a small blue galaxy, which is probably the optical counterpart of the detection FASHI\_1, and this is corroborated by the fact that FASHI\_1 shows a weaker emission (lower linewidths and lower peak flux density). The offset in right ascension is almost the same for the two \ion{H}{i} detections with respect to the potential optical counterparts.

\subsection{The case of PGC 001109}\label{subsec:pgc001109}
The lower panel in Figure \ref{fig:issues_catalogue_positionOffset} shows the \ion{H}{i} detection near PGC 001109.
This case is slightly different from the previous ones, since HIPASS did not detect the galaxy, but some \ion{H}{i} properties are listed in the work by \cite{1998A&AS..130..333T}, and are reported in the last section of the Table \ref{tab:fastIssues}. The centre of the FASHI detection is slightly offset with respect to PGC 001109, but the galaxy is partially within the 3\arcmin\ beam; however, the velocity listed in the FASHI catalogue is significantly different from the radial velocity reported by \cite{1998A&AS..130..333T}. Taking into account the quite significant differences between the two radio telescopes used by the FASHI collaboration and by Theureau et al. (who used the meridian-transit Nançay radio telescope with an equivalent aperture of 94 meters), the W$_{50}$ and the W$_{20}$ values are compatible. Again, the velocity offset is $\sim 2000$ km s$^{-1}$, as in the previous cases.


\bsp	
\label{lastpage}
\end{document}